\documentclass[12pt]{article}
\usepackage{amscd}
\usepackage{bbm}
\usepackage{mathrsfs}
\usepackage{amssymb}
\usepackage{graphics}
\usepackage{graphicx}
\usepackage{amsfonts}
\usepackage{amsmath}
\usepackage{caption}
\usepackage{titlesec}
\usepackage{subcaption}
\usepackage{float}
\usepackage[numbers,sort&compress]{natbib}
\titleformat{\section}{\large\bfseries}{\thesection. }{0em}{}
\titleformat{\subsection}{\normalsize\bfseries}{\thesubsection. }{0em}{}
\numberwithin{equation}{section}
\begin{document}
	\date{}
	\title{The Riemann–Hilbert approach for the nonlocal derivative nonlinear Schr\"{o}dinger equation with nonzero boundary conditions}
	
	\author{Xin-Yu Liu, Rui Guo$ \thanks{Corresponding author,
			gr81@sina.com}$ \
		\\
		\\{\em
			School of Mathematics, Taiyuan University of} \\
		{\em Technology, Taiyuan 030024, China} } \maketitle
	
	\begin{abstract}		
		In this paper, the nonlocal reverse space-time derivative nonlinear Schr\"{o}dinger equation under nonzero boundary conditions is investigated using the Riemann-Hilbert (RH) approach. The direct scattering problem focuses on the analyticity, symmetries, and asymptotic behaviors of the Jost eigenfunctions and scattering matrix functions, leading to the construction of the corresponding RH problem. Then, in the inverse scattering problem, the Plemelj formula is employed to solve the RH problem. So the reconstruction formula, trace formulae, $\theta$ condition, and exact expression of the single-pole and double-pole solutions are obtained. Furthermore, dark-dark solitons, bright-dark solitons, and breather solutions of the reverse space-time derivative nonlinear Schr\"{o}dinger equation are presented along with their dynamic behaviors summarized through graphical simulation.
		
		\vspace{7mm}\noindent\emph{Keywords}: Riemann–Hilbert approach; Reverse space-time derivative nonlinear Schr\"{o}dinger equation; Nonzero boundary conditions
	\end{abstract}

	\newpage
	\section{Introduction}
	\hspace{0.7cm}Nonlinear evolution equations can explain a wide range of complex nonlinear phenomena in nature, making a crucial research focus within the field of nonlinear science \cite{g1,g2}. Therefore, it is imperative to solve the nonlinear evolution equations and obtain exact solutions. Various effective methods for solving soliton solutions, breather solutions, and rogue wave solutions of nonlinear evolution equations have been developed, including Darboux transformation \cite{g3,g4}, Hirota bilinear method \cite{g5,g6}, inverse scattering transformation (IST) \cite{g7,g8,g9,g10}, and RH method \cite{g11,g12,g13,g14}, etc. Among these methods, RH method is a new form of IST, which can be applied in diverse scenarios without the Gel'fand-Levitan-Marchenko equation, and is more versatile than IST. The RH method can effectively solve a broad class of mathematical problems and has emerged as a prominent area of research within the field of mathematics. Building upon this foundation, exact solutions for numerous local nonlinear evolution equations are constructed and analyzed.
	
	Since Ablowitz introduced a new nonlinear integrable equation in 2013, namely the new nonlocal nonlinear Schr\"{o}dinger (NLS) equation \cite{g15}, nonlocal systems have gained attention. After then, Ablowitz utilized IST to conduct a comprehensive analysis of the Cauchy problem with rapidly decaying initial data for the nonlocal NLS equations and the solution under nonzero boundary conditions in 2016 and 2017, respectively \cite{g16,g17}. In addition to the nonlocal NLS equations, other nonlocal integrable equations have also been introduced and extensively studied, such as the nonlocal Hirota equation \cite{g18,g19}, the nonlocal Lakshmanan-Porsezian-Daniel equation \cite{g20}, the nonlocal modified KdV equation \cite{g21,g22}, the nonlocal sine-Gorden equation \cite{g23}, etc. These nonlocal equations give rise to numerous new physical phenomena and hold significant physical implications.
	
	In addition, the derivative NLS equation 
	\begin{equation}
		iq_t+q_{xx}+i\sigma \left( |q|^2q \right)_{x}=0,\ \sigma =\pm 1
	\end{equation}
	has been extensively investigated \cite{g24,g25,g26,g27}, which is of wide applications in plasma physics \cite{g28}. By adding a nonlocal term, the following two classes of nonlocal derivative NLS equations can be derived \cite{g29}:
	
	(1) Parity-time derivative NLS (PT-DNLS) equation:
	\begin{equation}
		iq_t\left( x,t \right) +q_{xx}\left( x,t \right) +\sigma \left( q^2\left( x,t \right) q^*\left( -x,t \right) \right) _x=0,  \,\sigma =\pm 1,\ 
	\end{equation}
	
	(2) Reverse space-time derivative NLS (RST-DNLS) equation:
	\begin{equation}\label{e1}
		q_t\left( x,t \right) =iq_{xx}\left( x,t \right) +\sigma \left( q^2\left( x,t \right) q\left( -x,-t \right) \right) _x,\,\sigma =\pm 1,\\
	\end{equation}
	where the subscripts denote the partial derivatives, the asterisk * represents complex conjugation, $\sigma =-1$ and $\sigma =1$ indicate the focusing and defocusing cases, respectively. 
	
	The above nonlocal derivative NLS equations play a pivotal role in the investigation of wave propagation in the fields of nonlinear optics, plasma physics, and ocean water waves \cite{g30,g31}. Soliton, breather, and rogue wave have been generated for nonlocal derivative NLS equations by employing Darboux transformation \cite{g32,g33}. The RH method has also been used to study the nonlocal derivative NLS equations for zero boundary cases \cite{g34,g35,g36}. The nonlocal Gerdjikov-Ivanov equation under nonzero boundary conditions was investigated in \cite{g37}, yielding only soliton solutions and no breather solutions. However, to our knowledge, the study of the RST-DNLS equation under nonzero boundary conditions is yet unexplored. Meanwhile, the research under nonzero boundary conditions is more intricate compared to that under zero boundary conditions. To further explore the relationship between $(x,t)$ and $(-x,-t)$ and obtain the breather solutions, we will employ the RH method to investigate Eq.~(\ref{e1}) under the nonzero boundary conditions
	\begin{equation}\label{e6}
		q\left( x,t \right) \rightarrow q_{\pm},\ x\rightarrow \pm \infty ,\\
	\end{equation}
	where\ $ |q_{\pm}|=q_{0}>0 $ and $q_{\pm}=\eta q_{\mp}^{\ast},\ \eta =\pm 1$, such that the branch points of the eigenvalues are on the real or imaginary axis. The nonlocal RST-DNLS equation possesses the compatible system as follows \cite{g38}:
	\begin{align}\label{e2}
		\varPsi _x&=X\varPsi , X\left( x,t,\lambda \right) =-i\lambda ^2\sigma _3+\lambda Q ,\\\label{e3}
		\varPsi _t&=T\varPsi ,\ T\left( x,t,\lambda \right) =-2i\lambda ^4\sigma _3+2\lambda ^3Q-i\lambda ^2Q^2-i\lambda Q_x+\lambda Q^3 , 
	\end{align}
	where $\varPsi=\varPsi(x,t;\lambda)$ is a $2\times2$ matrix eigenfunction, $\lambda$ is the complex spectral parameter, and
	\begin{subequations}
		\begin{align}
			\sigma _3=\left( \begin{matrix}
				1&		0\\
				0&		-1\\
			\end{matrix} \right),\ 
			Q=\left( \begin{matrix}
				0&		q(x,t)\\
				\sigma q(-x,-t)&		0\\
			\end{matrix} \right) . \nonumber
		\end{align}
	\end{subequations}
	
	The outline of this paper will be organized as follows: In Section 2, we focus on the direct scattering problem, analyze the analytic, symmetric, and asymptotic properties of Jost eigenfunctions and scattering matrix functions, as well as construct the corresponding RH problem. In Section 3, the inverse scattering problem for single-pole case will be discussed. We can obtain reconstruction formula and give the exact solution expression of the RH problem associated with the equation. Finally, the dynamic and propagation characteristics of soliton solutions with different parameters are analyzed. In Section 4, we discuss the inverse scattering problem in the double-pole case and obtain the exact expression of the solution in this scenario. The conclusion is presented in Section 5.

	\section{Direct Scattering problem}
	\hspace{0.7cm}The properties of eigenfunctions and scattering matrix functions are discussed in this section, including their analyticity, symmetries, and asymptotic behaviors. Relevant RH problem is presented by combining these properties. The analysis process is complicated by the appearance of Riemann surface, so the uniformization variable is introduced to simplify it.
	\subsection{Jost Solutions, Analyticity, and Continuity}
	\hspace{0.7cm}When $x\rightarrow \pm \infty$, according to the boundary conditions (\ref{e6}), the asymptotic scattering problem can be given by
	\begin{align}\label{e4}
		\varPsi _x&=X_{\pm}\varPsi ,\   X_{\pm}\left( x,t,\lambda \right) =-i\lambda ^2\sigma _3+\lambda Q_{\pm},\\\label{e5}
		\varPsi _t&=T_{\pm}\varPsi ,\ \  T_{\pm}\left( x,t,\lambda \right) =\left( 2\lambda ^2+\sigma \eta q_0^2 \right) X_{\pm},
	\end{align}
	where 
	\begin{align}
		Q_{\pm}=\left( \begin{matrix}
			0&		q_{\pm}\\
			\sigma q_{\mp}&		0\\
		\end{matrix} \right) =\left( \begin{matrix}
			0&		q_{\pm}\\
			\sigma \eta q_{\pm}^{\ast}&		0\\
		\end{matrix} \right) .
	\end{align}
	It can be easily calculated that the eigenvalues of the matrix $X_{\pm}$ are $\pm i\lambda\sqrt{\lambda^2-\sigma\eta q_{0}^2}$. Clearly, the eigenvalues are double-branched. Hence, we introduce a two-sheeted Riemann surface defined as follows:
	\begin{equation}
		k ^2=\lambda^2-\sigma\eta q_{0}^{2}  .
	\end{equation} 
	\hspace{0.7cm}With the above definition, we find the branch points are $\pm q_{0}$ when $\sigma\eta = 1$, and the branch points are $\pm iq_{0}$ when $\sigma\eta =-1$. Introducing the uniformization variable $z=k+\lambda $ results in two single-valued functions:
	\begin{equation}
		k=\frac{1}{2}\left( z-\sigma \eta \frac{q_0^2}{z} \right), \ 
		\lambda =\frac{1}{2}\left( z+\sigma \eta \frac{q_0^2}{z} \right) . 
	\end{equation}
	Through analysis, the mapping between two-sheet Riemann surface and complex $z$ plane can be obtained. In view of this, we can proceed with the discussion of the scattering problem in the complex $z$ plane (see Fig. 1). 
	\begin{figure}[H]
		\centering
		\begin{minipage}[c]{0.45\textwidth} 
			\centering
			\includegraphics[height=6cm,width=\textwidth]{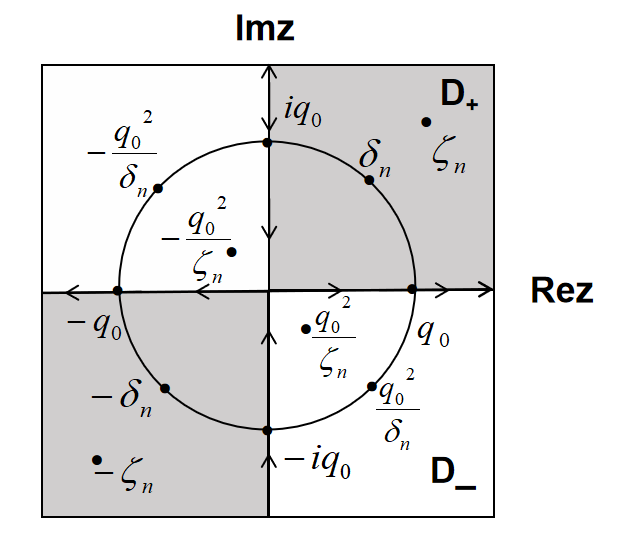} 
		\end{minipage}
		\caption{Complex $z$-plane, showing the region $D_+$(grey region), the region $D_-$(white region), the distribution of discrete spectrum, and the contours for the RH problem.}
	\end{figure}
	For convenience, let
	$$
	D_+=\left\{ z\in \mathbb{C}|\text{Re}z\text{Im}z>0 \right\} ,\ D_-=\left\{ z\in \mathbb{C}|\text{Re}z\text{Im}z<0 \right\} ,\ \Sigma =\mathbb{R}\cup i\mathbb{R}\setminus \left\{ 0 \right\} ,\ 
	$$
	which imply that $\text{Im}\left( \lambda \left( z \right) k\left( z \right) \right) >0$ when $z\in D_+$, and $\text{Im}\left( \lambda \left( z \right) k\left( z \right) \right) <0$ when $z\in D_-$. Now, $X_{\pm}$ and $T_{\pm}$ can be converted into diagonal matrices, i.e.,
	\begin{equation}
		X_{\pm}=M_{\pm}\left( ik\lambda \sigma _3 \right) M_{\pm}^{-1},\ T_{\pm}=M_{\pm}\left( ik\lambda \left( 2\lambda ^2+\sigma \eta q_0^2 \right) \sigma _3 \right) M_{\pm}^{-1} , 
	\end{equation}
	where
	\begin{equation}
		M_{\pm}=\left( \begin{matrix}
			\frac{-iq_{\pm}}{z}&		1\\
			1&		\frac{i\sigma \eta q_{\pm}^{\ast}}{z}\\
		\end{matrix} \right) . 
	\end{equation}
	
	Thus, we get the Jost eigenfunctions $\varPsi _{\pm}\left( x,t,z \right) $ that satisfies the two parts of Lax pair (\ref{e2})-(\ref{e3}) under the boundary conditions
	\begin{equation}
		\varPsi _{\pm}= M_{\pm}e^{i\theta \left( z \right) \sigma _3}, \   x\rightarrow \pm \infty , 
	\end{equation}
	where $\theta \left( z \right) =k\left( z \right) \lambda \left( z \right) \left[ x+\left( 2 \lambda^2\left( z \right) +\sigma \eta q_{0}^{2} \right) t \right]$.
	
	Next, the modified Jost eigenfunctions $\nu _{\pm}\left( x,t,z \right) $ can be taken as 
	\begin{equation}\label{e7}
		\nu _{\pm}=\varPsi _{\pm}e^{-i\theta\left( z \right)  \sigma _3}.
	\end{equation} 
	As a result, $\nu _{\pm}\rightarrow M_{\pm}\ $as$\ x\rightarrow \pm \infty $. Combining with the properties of differential functions, the modified Jost eigenfunctions $\nu _{\pm}\left( x,t,z \right) $ subject to the following linear integral equation:
	\begin{subequations}
		\begin{align}
			&  \nu _-\left( x,t,z \right) =M_-+\lambda \int_{-\infty}^x{M_-e}^{ik\lambda \left( x-y \right) \hat{\sigma}_3}\left[ M_{-}^{-1}\varDelta Q_-\left( y,t,z \right) \nu _-\left( y,t,z \right) \right] dy,\\
			&\nu _+\left( x,t,z \right) =M_+-\lambda \int_x^{\infty}{M_+e}^{ik\lambda \left( x-y \right) \hat{\sigma}_3}\left[ M_{+}^{-1}\varDelta Q_+\left( y,t,z \right) \nu _+\left( y,t,z \right) \right] dy.
		\end{align}
	\end{subequations}
	
	Since the Jost solutions $\varPsi _{\pm}\left( x,t,z \right) $ are the fundamental matrix solutions of Eqs.~(\ref{e2})-(\ref{e3}), so there exist a constant scattering matrix $ U(z) $ such that
	\begin{equation}\label{e8}
		\nu _+\left( x,t,z \right) =\nu _-\left( x,t,z \right) e^{i\theta\left( z \right)  \hat{\sigma}_3}U\left( z \right) ,
	\end{equation}
	where
	$$
	U\left( z \right) =\left( \begin{matrix}
		u_{11}\left( z \right)&		u_{12}\left( z \right)\\
		u_{21}\left( z \right)&		u_{22}\left( z \right)\\
	\end{matrix} \right) .
	$$
	
	Let $\nu _{\pm}=\left( \nu _{\pm,1},\nu _{\pm,2} \right)$, the subscripts 1 and 2 denote the columns of a 2 × 2 matrix. Similar to Ref.~[13], the analytic properties of the eigenfunctions can be derived by constructing Neumann series, i.e., when $
	q-q_{\pm}\in L^1\left( \mathbb{R} \right) $, $\nu _{+,1} , \ \nu _{-,2}$ can be analytically extended to $D_{+}$, $\nu _{+,2} ,\ \nu _{-,1}$ can be analytically extended to $D_{-}$ . Hence, we can rewrite $\nu _{\pm}$ as 
	$$
	\nu _+=\left( \nu _{+,1}^{+}, \nu _{+,2}^{-} \right) ,\  \nu _-=\left( \nu _{-,1}^{-},\nu _{-,2}^{+} \right) . 
	$$
	
	In terms of Lax pair (\ref{e2})-(\ref{e3}), $tr(X)=tr(T)=0$. Then, using Abel formula and transformation (\ref{e7}), we have
	$$
	\det \left( \nu _{\pm} \right) _x=\det \left( \nu _{\pm} \right) _t=0 , 
	$$
	then
	$$
	\det \left( \nu _{\pm} \right) =\underset{x\rightarrow \infty}{\lim}\det \left( \nu _{\pm} \right) =\det \left( M_{\pm} \right) =\tau ,
	$$
	where $\tau=\frac{\sigma \eta q_{0}^{2}}{z^2}-1 $. 
	
	Moreover, the scattering coefficient $u_{11}\left( z \right),\ u_{12}\left( z \right),\ u_{21}\left( z \right),\ u_{22}\left( z \right)$ can be expressed via eigenfunctions,
	\begin{subequations}
		\begin{align}\label{e16}
			u_{11}\left( z \right) &=\frac{\det \left( \nu _{+,1},\nu _{-,2} \right)}{\tau}\,,\,\,\ \ \ \ \ \ \ \,\,u_{21}\left( z \right) =\frac{e^{2i\theta}\det \left( \nu _{-,1},\nu _{+,1} \right)}{\tau}, \\\label{e17}
			u_{12}\left( z \right) &=\frac{e^{-2i\theta}\det \left( \nu _{+,2},\nu _{-,2} \right)}{\tau},\,\,\,\,\,\,u_{22}\left( z \right) =\frac{\det \left( \nu _{-,1},\nu _{+,2} \right)}{\tau}.		
		\end{align}
	\end{subequations}
	As can be seen from the above, suppose $q-q_{\pm}\in L^1\left( \mathbb{R} \right) $, $u_{11}(z)$ can be analytically extended to $D_{+}$, $u_{22}(z)$ can be analytically extended to $D_{-}$, $u_{12}(z)$ and $u_{21}(z)$ are continuous in $\Sigma $. Finally, in order to facilitate discussion of RH problem, we define the reflection coefficient as follows:
	$$
	\rho \left( z \right) =\frac{u_{21}(z)}{u_{11}(z)},\,\,\tilde{\rho}\left( z \right) =\frac{u_{12}(z)}{u_{22}(z)}.
	$$
	
	\subsection{Symmetries of eigenfunctions and scattering coefficients}
	\hspace{0.7cm}This section examines the symmetries of eigenfunctions and scattering matrix functions, which are crucial for analyzing discrete spectrum and residual conditions in the inverse scattering problem. Unlike the local case, the nonlocal case involves three symmetries: firstly, $z\rightarrow -z$, implying $\left( k,\lambda \right) \rightarrow \left( -k,-\lambda \right)$; secondly, $z\rightarrow \sigma \eta \frac{q_{0}^{2}}{z}$, implying $ \left( k,\lambda \right) \rightarrow \left( k,-\lambda \right) $; and finally, by combining the transformation of $\left( x,t \right) \rightarrow \left( -x,-t \right) $, the following three symmetries of the eigenfunctions and the scattering matrix functions are obtained.
	\begin{align}
		&\varPsi _{\pm}\left( x,t,z \right) =\left\{ \begin{array}{l}
			\sigma _1\varPsi _{\mp}\left( -x,-t,-z \right) \sigma _1\ ,\ \ \ \ \sigma =1\,\,,\\
			-\sigma _2\varPsi _{\mp}\left( -x,-t,-z \right) \sigma _2\ ,\,\ \,\sigma =-1\,\,,\\
		\end{array} \right.   \\
		&\varPsi _{\pm}\left( x,t,z \right) =\frac{i}{z}\varPsi _{\pm}\left( x,t,\sigma \eta \frac{q_{0}^{2}}{z} \right) \sigma _1Q_{\pm}\sigma _{\ast} ,\\
		&\varPsi _{\pm}\left( x,t,z \right) =-\sigma _3\varPsi _{\pm}\left( x,t,-z \right) \sigma _3 . 
	\end{align}
	Then, expand the above symmetries in columns,
	\begin{subequations}\label{e15}
		\begin{align}
			&\varPsi _{\pm ,1}\left( x,t,z \right) =\left\{ \begin{array}{l}
				\sigma _1\varPsi _{\mp ,2}\left( -x,-t,-z \right) \,,\qquad\sigma =1\,,\\
				-i\sigma _2\varPsi _{\mp ,2}\left( -x,-t,-z \right)\ ,\,\ \ \sigma =-1\,\,,\\
			\end{array} \right. \\
			&\varPsi _{\pm ,1}\left( x,t,z \right) =-\sigma _3\varPsi _{\pm ,1}\left( x,t,-z \right) ,\qquad\ 
			\ \varPsi _{\pm ,2}\left( x,t,z \right) =\sigma _3\varPsi _{\pm ,2}\left( x,t,-z \right) ,\\ \label{e21}
			&\varPsi _{\pm ,1}\left( x,t,z \right) =-\frac{iq_{\pm}}{z}\varPsi _{\pm ,2}\left( x,t,\frac{\sigma \eta q_0^2}{z} \right) ,\ \ \varPsi _{\pm ,2}\left( x,t,z \right) =\frac{i\sigma \eta q_{\pm}^{\ast}}{z}\varPsi _{\pm ,1}\left( x,t,\frac{\sigma \eta q_0^2}{z} \right) .\ \ 
		\end{align}
	\end{subequations}
	In what follows, combining the above symmetries of eigenfunctions together with the relation (\ref{e8}) we have
	\begin{align}
		&U\left( z \right) =\left\{ \begin{array}{l}
			\sigma _1\left( U\left( -z \right) \right) ^{-1}\sigma _1\,\,,\,\,\sigma =1\,\,,\\
			\text{}\sigma _2\left( U\left( -z \right) \right) ^{-1}\sigma _2\,\,,\,\,\sigma =-1\,\,,\text{}\\
		\end{array} \right.\\
		&	U\left( z \right) =\left( \sigma _1Q_-\sigma _{\ast} \right) ^{-1}U\left( \sigma \eta \frac{q_{0}^{2}}{z} \right) \sigma _1Q_+\sigma _{\ast}\ ,\\
		&U\left( z \right) =\sigma _3U\left( -z \right) \sigma _3\ ,
	\end{align}
	where
	\begin{equation}
		\sigma _1=\left( \begin{matrix}
			0&		1\\
			1&		0\\
		\end{matrix} \right) ,\ \sigma _2=\left( \begin{matrix}
			0&		-i\\
			i&		0\\
		\end{matrix} \right) ,\ \sigma _{\ast}=\left( \begin{matrix}
			0&		1\\
			-1&		0\\
		\end{matrix} \right) .
	\end{equation}
	Or in component form,
	\begin{subequations}\label{e14}
		\begin{align}
			&u_{11}\left( z \right) =u_{11}\left( -z \right) ,\qquad\ \ \ u_{22}\left( z \right) =u_{22}\left( -z \right) ,\\
			&u_{11}\left( z \right) =\frac{q_+}{q_-}u_{22}\left( \frac{\sigma \eta q_0^2}{z} \right) ,\ u_{22}\left( z \right) =\frac{q_-}{q_+}u_{11}\left( \frac{\sigma \eta q_0^2}{z} \right) .
		\end{align}
	\end{subequations}

	\subsection{Asymptotic behaviors of eigenfunctions and scattering coefficients}
	\hspace{0.7cm}In order to propose and solve the RH problem in the following section, it is essential to provide the asymptotic behaviors of the Jost eigenfunctions and the scattering matrix functions as $z\rightarrow \infty$ and $z\rightarrow 0$. The asymptotic properties of the Jost eigenfunctions and scattering matrix functions are as follows:
	
	\begin{equation}
		\nu _{\pm}\left( x,t,z \right) =e^{im_{\pm}\sigma _{\ast}}+O\left( \frac{1}{z} \right) ,\ z\rightarrow \infty , 
	\end{equation}
	\begin{equation}
		\nu _{\pm}\left( x,t,z \right) =\frac{i}{z}e^{im_{\pm}\sigma _{\ast}}\sigma \eta \sigma _3Q_{\pm}^{\ast}+O\left( 1 \right) ,\ z\rightarrow 0 , 
	\end{equation}
	where
	\begin{equation}
		m_{\pm}=-\frac{1}{2}\int\limits_{\pm \infty}^x{\left( \sigma \eta q_{0}^{2}-\sigma q\left( y,t \right) q\left( -y,-t \right) \right) dy}. 
	\end{equation}
	Proof:
	
	The modified Jost eigenfunctions $\nu _{\pm}$ has the following expansion as $z\rightarrow \infty$,
	\begin{equation}\label{e9}
		\nu _{\pm}=\ \nu _{\pm}^{\left( 0 \right)}\left( x,t \right) +\frac{\nu _{\pm}^{\left( 1 \right)}\left( x,t \right)}{z}+\frac{\nu _{\pm}^{\left( 2 \right)}\left( x,t \right)}{z^2}+\cdot \cdot \cdot \ ,\ z\rightarrow \infty .\
	\end{equation}
	Substituting Eqs.~(\ref{e7}) and (\ref{e9}) into Eq.~(\ref{e2}), we find the following equation:
	\begin{equation}
		\begin{aligned}
			\left( \nu _{\pm}^{\left( 0 \right)}\left( x,t \right) +\frac{\nu _{\pm}^{\left( 1 \right)}\left( x,t \right)}{z}+\frac{\nu _{\pm}^{\left( 2 \right)}\left( x,t \right)}{z^2} \right) _x&=\left( -i\lambda ^2\sigma _3+\lambda Q_{\pm} \right) \left( \nu _{\pm}^{\left( 0 \right)}\left( x,t \right) +\frac{\nu _{\pm}^{\left( 1 \right)}\left( x,t \right)}{z}+\frac{\nu _{\pm}^{\left( 2 \right)}\left( x,t \right)}{z^2} \right) \\
			&+\lambda \varDelta Q_{\pm}\left( \nu _{\pm}^{\left( 0 \right)}\left( x,t \right) +\frac{\nu _{\pm}^{\left( 1 \right)}\left( x,t \right)}{z}+\frac{\nu _{\pm}^{\left( 2 \right)}\left( x,t \right)}{z^2} \right) \\ 
			&-i\left( \nu _{\pm}^{\left( 0 \right)}\left( x,t \right) +\frac{\nu _{\pm}^{\left( 1 \right)}\left( x,t \right)}{z}+\frac{\nu _{\pm}^{\left( 2 \right)}\left( x,t \right)}{z^2} \right) k\lambda \sigma _3 .
		\end{aligned}
	\end{equation}
	Comparing the power coefficient of $z$ in the above equation, we derive
	\begin{align}\label{e10}
		&O\left( z^2 \right) :\ -\frac{i}{4}\sigma _3\nu _{\pm}^{\left( 0 \right)}-\frac{i}{4}\nu _{\pm}^{\left( 0 \right)}\sigma _3=0 ,\\ \label{e11}
		&O\left( z \right) :-\frac{i}{4}\sigma _3\nu _{\pm}^{\left( 1 \right)}+\frac{1}{2}Q_{\pm}\nu _{\pm}^{\left( 0 \right)}-\frac{i}{4}\nu _{\pm}^{\left( 1 \right)}\sigma _3+\frac{1}{2}\varDelta Q_{\pm}\nu _{\pm}^{\left( 0 \right)}=0 , \\ \label{e12}
		&O\left( 1 \right) :\nu _{\pm ,x}^{\left( 0 \right)}=-\frac{i}{4}\sigma _3\nu _{\pm}^{\left( 2 \right)}-\frac{i}{2}\sigma \eta q_{0}^{2}\sigma _3\nu _{\pm}^{\left( 0 \right)}+\frac{1}{2}Q_{\pm}\nu _{\pm}^{\left( 1 \right)}-\frac{i}{4}\nu _{\pm}^{\left( 2 \right)}\sigma _3+\frac{1}{2}\varDelta Q_{\pm}\nu _{\pm}^{\left( 1 \right)} .
	\end{align}
	Now, we assume 
	$$
	\nu _{\pm}^{\left( 0 \right)}\left( x,t \right) =\left( \begin{matrix}
		\nu _{\pm ,11}^{\left( 0 \right)}&		\nu _{\pm ,12}^{\left( 0 \right)}\\
		\nu _{\pm ,21}^{\left( 0 \right)}&		\nu _{\pm ,22}^{\left( 0 \right)}\\
	\end{matrix} \right) ,\ \nu _{\pm}^{\left( 1 \right)}\left( x,t \right) =\left( \begin{matrix}
		\nu _{\pm ,11}^{\left( 1 \right)}&		\nu _{\pm ,12}^{\left( 1 \right)}\\
		\nu _{\pm ,21}^{\left( 1 \right)}&		\nu _{\pm ,22}^{\left( 1 \right)}\\
	\end{matrix} \right) .
	$$
	Indeed, from Eq.~(\ref{e11}), the potential function $q(x,t)$ and $q(-x,-t)$ can be given by
	\begin{equation}
		q\left( x,t \right) =i\nu _{\pm ,11}^{\left( 1 \right)}(x,t)\left( \nu _{\pm ,21}^{\left( 0 \right)}(x,t) \right) ^{-1},\ q\left( -x,-t \right) =-i\sigma \nu _{\pm ,22}^{\left( 1 \right) }(x,t) \left( \nu _{\pm ,12}^{\left( 0 \right)}(x,t) \right) ^{-1} .
	\end{equation}
	Then, from Eqs.~(\ref{e10}) and (\ref{e12}), we observe that $\nu _{\pm}^{\left( 0 \right)}(x,t)$ is an off diagonal matrix and $\nu _{\pm}^{\left( 0 \right)}(x,t)$ can be expressed as
	$$
	\nu _{\pm}^{\left( 0 \right)}\left( x,t \right) =Pe^{-\frac{i}{2}\int\limits_{\pm \infty}^x{\left( \sigma \eta q_{0}^{2}-\sigma q\left( y,t \right) q\left( -y,-t \right) \right) dy}\sigma _{\ast}},
	$$
	where $P$ is a diagonal matrix, combining that $ \nu _{\pm}^{\left( 0 \right)}=P\sigma _1 $ as $ x\rightarrow \pm \infty$, thus
	$$
	P\sigma _1=\underset{x\rightarrow \pm \infty}{\lim}\underset{z\rightarrow \infty}{\lim}\nu _{\pm}\left( x,t,z \right) =\underset{z\rightarrow \infty}{\lim}M_{\pm}\left( z \right) =\underset{z\rightarrow \infty}{\lim}\left( \sigma _1+\frac{i}{z}Q_{\pm}\sigma _{\ast} \right) =\sigma _1 , 
	$$
	i.e.,
	$$
	\nu _{\pm}\left( x,t,z \right) =e^{im_{\pm}\sigma _{\ast}}+O\left( \frac{1}{z} \right) .
	$$
	
	In what follows, expand the modified Jost eigenfunctions $ \nu _{\pm} $ as $z\rightarrow 0$,
	\begin{equation}\label{e13}
		\nu _{\pm}=\ \frac{\nu _{\pm}^{\left( -1 \right)}\left( x,t \right)}{z}+\nu _{\pm}^{\left( 0 \right)}\left( x,t \right) +z\nu _{\pm}^{\left( 1 \right)}\left( x,t \right) +\cdot \cdot \cdot,\ z\rightarrow 0 .
	\end{equation}
	Similarly, substituting Eqs.~(\ref{e7}) and (\ref{e13}) into Eq.~(\ref{e2}), we find the following equation:
	\begin{equation}
		\begin{aligned}
			\left( \frac{\nu _{\pm}^{\left( -1 \right)}\left( x,t \right)}{z}+\nu _{\pm}^{\left( 0 \right)}\left( x,t \right) +z\nu _{\pm}^{\left( 1 \right)}\left( x,t \right) \right) _x&=\left( -i\lambda ^2\sigma _3+\lambda Q_{\pm} \right) \left( \frac{\nu _{\pm}^{\left( -1 \right)}\left( x,t \right)}{z}+\nu _{\pm}^{\left( 0 \right)}\left( x,t \right) +z\nu _{\pm}^{\left( 1 \right)}\left( x,t \right) \right) \\
			&+\lambda \varDelta Q_{\pm}\left( \frac{\nu _{\pm}^{\left( -1 \right)}\left( x,t \right)}{z}+\nu _{\pm}^{\left( 0 \right)}\left( x,t \right) +z\nu _{\pm}^{\left( 1 \right)}\left( x,t \right) \right) \\
			&-i\left( \frac{\nu _{\pm}^{\left( -1 \right)}\left( x,t \right)}{z}+\nu _{\pm}^{\left( 0 \right)}\left( x,t \right) +z\nu _{\pm}^{\left( 1 \right)}\left( x,t \right) \right) k\lambda \sigma _3 . 
		\end{aligned}
	\end{equation}
	Comparing the power coefficient of $z$ in the above equation, we derive
	\begin{align}\label{14}
		&O\left( z^{-3} \right) :\ -\frac{i}{4}q_0^4\sigma _3\nu _{\pm}^{\left( -1 \right)}+\frac{i}{4}\nu _{\pm}^{\left( -1 \right)}q_0^4\sigma _3=0 ,\\\label{15}
		&O\left( z^{-2} \right) :-\frac{i}{4}q_0^4\sigma _3\nu _{\pm}^{\left( 0 \right)}+\frac{1}{2}\sigma \eta q_0^2Q_{\pm}\nu _{\pm}^{\left( -1 \right)}+\frac{i}{4}\nu _{\pm}^{\left( 0 \right)}q_0^4\sigma _3+\frac{1}{2}\sigma \eta q_0^2\varDelta Q_{\pm}\nu _{\pm}^{\left( -1 \right)}=0 ,\\\label{16}
		&O\left( z^{-1} \right) :\nu _{\pm ,x}^{\left( -2 \right)}=-\frac{i}{4}q_{0}^{4}\sigma _3\nu _{\pm}^{\left( 1 \right)}-\frac{i}{2}\sigma \eta q_{0}^{2}\sigma _3\nu _{\pm}^{\left( -1 \right)}+\frac{i}{4}\nu _{\pm}^{\left( 1 \right)}q_{0}^{4}\sigma _3+\frac{1}{2}\sigma \eta q_{0}^{2}Q\nu _{\pm}^{\left( 0 \right)} .
	\end{align}
	Now, we assume 
	$$
	\nu _{\pm}^{\left( 0 \right)}\left( x,t \right) =\left( \begin{matrix}
		\nu _{\pm ,11}^{\left( 0 \right)}&		\nu _{\pm ,12}^{\left( 0 \right)}\\
		\nu _{\pm ,21}^{\left( 0 \right)}&		\nu _{\pm ,22}^{\left( 0 \right)}\\
	\end{matrix} \right) ,\ \nu _{\pm}^{\left( -1 \right)}\left( x,t \right) =\left( \begin{matrix}
		\nu _{\pm ,11}^{\left( -1 \right)}&		\nu _{\pm ,12}^{\left( -1 \right)}\\
		\nu _{\pm ,21}^{\left( -1 \right)}&		\nu _{\pm ,22}^{\left(-1 \right)}\\
	\end{matrix} \right) .
	$$
	Applying a similar analysis on Eq.~(\ref{15}) leads to
	$$
	\nu _{\pm ,12}^{\left( 0 \right)}=-i\sigma \eta \frac{q\left( x,t \right)}{q_{0}^{2}}\nu _{\pm ,22}^{\left( -1 \right)}\,\,,\,\,\nu _{\pm ,21}^{\left( 0 \right)}=i\eta \frac{q\left( -x,-t \right)}{q_{0}^{2}}\nu _{\pm ,11}^{\left( -1 \right)} .
	$$
	At the same time, from Eqs.~(\ref{14}) and (\ref{16}), we derive that $\nu _{\pm}^{\left( -1 \right)}\left( x,t \right)$ is a diagonal matrix and
	$$ 
	\nu _{\pm}^{\left( -1 \right)}\left( x,t \right) =\tilde{P}e^{\frac{i}{2}\int\limits_{\pm \infty}^x{\left( \sigma \eta q_{0}^{2}-\sigma q\left( y,t \right) q\left( -y,-t \right) \right) dy}\sigma _{\ast}},
	$$
	where $\tilde{P}$ is an off diagonal matrix, combining that $ \nu _{\pm}^{\left( -1 \right)}=\tilde{P}\sigma _1 $ as $ x\rightarrow \pm \infty $, thus
	$$
	\tilde{P}\sigma _1=\underset{z\rightarrow 0}{\lim}\underset{x\rightarrow \pm \infty}{\lim}z\nu _{\pm}\left( x,t,z \right) =\underset{z\rightarrow 0}{\lim}zM_{\pm}\left( z \right) =\underset{z\rightarrow 0}{\lim}\left( z\sigma _1+iQ_{\pm}\sigma _{\ast} \right) =iQ_{\pm}\sigma _{\ast} ,
	$$
	it follows that
	$$
	\tilde{P}=iQ_{\pm}\sigma _{\ast}\sigma _1=iQ_{\pm}\sigma _3 , 
	$$
	i.e.,
	$$
	\nu _{\pm}=\frac{iQ_{\pm}\sigma _3}{z}e^{-im_{\pm}\sigma _{\ast}}=\frac{i}{z}e^{im_{\pm}\sigma _{\ast}}\sigma \eta \sigma _3Q_{\pm}^{\ast}+O\left( z \right) . \qquad\Box 
	$$
	
	Then, the asymptotic properties of the scattering matrix are acquired from the relation (\ref{e8}),
	\begin{equation}
		U\left( z \right) =e^{i\bar{m}\sigma _3}+O\left( \frac{1}{z} \right),\ z\rightarrow \infty , 
	\end{equation}
	\begin{equation}
		U\left( z \right) =\left( \begin{matrix}
			\frac{q_+}{q_-}&		0\\
			0&		\frac{q_-}{q_+}\\
		\end{matrix} \right) e^{-i\bar{m}\sigma _3}+O\left( z \right) ,\ z\rightarrow 0 , 
	\end{equation}
	where 
	\begin{equation}
		\bar{m}=m_--m_+=\frac{1}{2}\int\limits_{-\infty}^{\infty}{\left( \sigma q\left( y,t \right) q\left( -y,-t \right) -\sigma \eta q_0^2 \right)}dy . 
	\end{equation}
	\section{Inverse scattering problem with single-pole}
	\subsection{Riemann-Hilbert problem}
	\hspace{0.7cm}Introducing two 2×2 matrices based on the modified Jost eigenfunctions,
	\begin{equation}\label{e35}
		W^+=\left( \frac{\nu _{+,1}}{u_{11}},\nu _{-,2} \right) ,\ W^-=\left( \nu _{-,1},\frac{\nu _{+,2}}{u_{22}} \right) ,
	\end{equation}
	then, from the properties of the modified Jost eigenfunctions and scattering coefficients obtained before, one has the following RH problem:\\
	(1) $W^\pm$ are meromorphic in $D_{\pm}\setminus \varSigma$ respectively, \\
	(2) Jump condition:
	\begin{equation}\label{e20}
		W^-\left( x,t,z \right) =W^+\left( x,t,z \right) \left( I-J\left( x,t,z \right) \right) , 
	\end{equation}
	(3) Asymptotic behaviors:
	\begin{subequations}
		\begin{align}
			&W^{\pm}\left( x,t,z \right) =e^{im_-\sigma _{\ast}}+O\left( \frac{1}{z} \right) ,\,z\rightarrow \infty ,\\
			&W^{\pm}\left( x,t,z \right) =\frac{i}{z}\sigma \eta e^{im_-\sigma _{\ast}}\sigma _3Q_-^{\ast}+O\left( z \right),\,z\rightarrow 0 ,
		\end{align}
	\end{subequations}
	where
	$$
	J=e^{i\theta \left( z \right) \hat{\sigma}_3}\left( \begin{matrix}
		0&		-\tilde{\rho}\left( z \right)\\
		\rho \left( z \right)&		\rho \left( z \right) \tilde{\rho}\left( z \right)\\
	\end{matrix} \right) . 
	$$
	
	Then, the solution of Eqs.~(\ref{e1})-(\ref{e6}) can be expressed by the solution of RH problem,
	\begin{equation}\label{e18}
		q\left( x,t \right) =ie^{im_{\pm}}\underset{z\rightarrow \infty}{\lim}\left( z\nu _{\pm} \right) _{11}=ie^{im_-}\underset{z\rightarrow \infty}{\lim}\left( zW^- \right) _{11} .
	\end{equation}
	
	\subsection{Discrete spectrum and residue conditions}
	\hspace{0.7cm}First, we assume that $u_{11} (z) $ has $N_1$ simple zeros in $D _{+}\bigcap\left\{ z\in \mathbb{C}\mid \text{Re}z>0,\text{Im}z>0,|z| \ne q_0 \right\} $, expressed as $\zeta_{n} , n=1,2,\cdot\cdot\cdot,N_1$, then $u_{11}\left( \zeta_n \right) =0,\,u_{11}^{'}\left( \zeta_n \right) \ne 0$. From the symmetry of the scattering matrix (\ref{e14}), if $u_{11}\left( \zeta_n \right) =0$, then $u_{22}\left( \frac{\sigma \eta q_{0}^{2}}{\zeta _n} \right) =u_{22}\left( -\frac{\sigma \eta q_{0}^{2}}{\zeta _n} \right) =u_{11}\left( -\zeta _n \right) =0 $. In addition, we suppose that $u_{11} (z) $ has $N_2$ simple zeros in $D _{+}\bigcap \left\{ z\in \mathbb{C}\mid \text{Re}z>0,\text{Im}z>0,|z| =q_0 \right\} $, expressed as $\delta_{n} , n=1,2,\cdot\cdot\cdot,N_2$, then $u_{22}\left( \frac{\sigma \eta q_{0}^{2}}{\delta _n} \right) =u_{22}\left( -\frac{\sigma \eta q_{0}^{2}}{\delta _n} \right) =u_{11}\left( -\delta _n \right) =0 $.  
	For convenience, we define
	\begin{equation}\label{e38}
		\xi _n=\left\{ \begin{array}{l}
			\zeta _n\,\,,\,\,\,\,\,\,\,\,\,\,\,\,\,\,\,\,\,\,\,\,\ \ n=1,\,\,2,\,\,\cdot \cdot \cdot \ N_1\,\,,\\
			-\zeta _{n-N_1}\,\,,\,\,\,\,\,\,\,\,\ \,n=N_1+1,\,\,N_1+2,\,\,\cdot \cdot \cdot \ 2N_1\,\,,\\
			\delta _{n-2N_1}\,\,\,\,,\,\,\,\,\,\,\,\,\,\,\ n=2N_1+1,\,\,2N_1+2,\,\,\cdot \cdot \cdot\ 2N_1+N_2\,\,,\\
			-\delta _{n-2N_1-N_2}\,\,,\,\,n=2N_1+N_2+1,\,\,2N_1+N_2+2,\,\,\cdot \cdot \cdot\ 2N_1+2N_2\,\,,\text{
			}\\
		\end{array}\,\, \right.
	\end{equation}
	and $\widehat{\xi }_n=\frac{\sigma \eta q_{0}^{2}}{\xi _n}\ ,\ $so we have $4N_1+4N_2 $ discrete spectra, thus the set of discrete spectra is
	$$
	Z=\left\{ \xi _n,\ \widehat{\xi }_n,\ n=1,\cdot \cdot \cdot ,\ 2N_1+2N_2 \right\} .
	$$
	\hspace{0.7cm}If $z\in Z\cap D_+$, then $u_{11}\left( z \right) =0$. According to Eq.~(\ref{e16}), $\varPsi _{+,1}\left( z \right)$ and $\varPsi _{-,2}\left( z \right)$ are linearly dependent. Similarly, if $z\in Z\cap D_-$, then $u_{22}\left( z \right) =0$. And in terms of Eq.~(\ref{e17}), $\varPsi _{+,2}\left( z \right)$ and $\varPsi _{-,1}\left( z \right)$ are also linearly dependent. For convenience, we let
	$$
	a\left( z \right) =\left\{ \begin{array}{l}
		\varPsi _{+,1}\left( z \right) /\varPsi _{-,2}\left( z \right) ,\,\,z\in Z\cap D_+ ,\\
		\varPsi _{+,2}\left( z \right) /\varPsi _{-,1}\left( z \right) ,\,\,z\in Z\cap D_- ,\\
	\end{array} \right. \ A\left( z \right) =\left\{ \begin{array}{l}
		a\left( z \right) /u_{11}^{'}\left( z \right) ,\,\,z\in Z\cap D_+ ,\\
		a\left( z \right) /u_{22}^{'}\left( z \right) ,\,\,z\in Z\cap D_- ,\\
	\end{array} \right. 
	$$
	taking into account the symmetries (\ref{e15}) of eigenfunctions, one has
	\begin{subequations}\label{e27}
		\begin{align}
			&a\left( x,t,z\right) a\left( -x,-t,-z \right) =\sigma, \\
			&a\left( x,t,z\right) =-a\left( x,t,-z \right), \\
			&a\left( x,t,z\right) =-\sigma a\left( x,t,\sigma \eta \frac{q_0^2}{z} \right). 		
		\end{align}
	\end{subequations}
	As a result, we have
	$$
	\underset{z=\xi _n}{\text{Res}}\,\,\left[ \frac{\nu _{+,1}\left( z \right)}{u_{11}\left( z \right)} \right] =\frac{\nu _{+,1}\left( \xi _n \right)}{u_{11}^{'}\left( \xi _n \right)}=\frac{a\left( \xi _n \right)}{u_{11}^{'}\left( \xi _n \right)}e^{-2i\theta \left( \xi _n \right)}\nu _{-,2}\left( \xi _n \right), 
	$$
	similarly, it can be concluded that
	$$
	\underset{z=\widehat{\xi }_n}{\text{Res}}\,\,\left[ \frac{\nu _{+,2}\left( z \right)}{u_{22}\left( z \right)} \right] =\frac{\nu _{+,2}\left( \widehat{\xi }_n \right)}{u_{22}^{'}\left( \widehat{\xi }_n \right)}=\frac{a\left( \widehat{\xi }_n \right)}{u_{22}^{'}\left( \widehat{\xi }_n \right)}e^{2i\theta \left( \widehat{\xi }_n \right)}\nu _{-,1}\left( \widehat{\xi }_n \right) .
	$$
	Furthermore, the residual conditions of $W^+$and $W^-$ can be taken as
	\begin{align}
		&\underset{z=\xi _n}{\text{Res}}\,\,W^+\left( z \right) =\left( A\left( \xi _n \right) e^{-2i\theta \left( \xi _n \right)}\nu _{-,2}\left( x,t,\xi _n \right), \ 0 \right) ,\ n=1,\cdot \cdot \cdot ,\ 2N_1+2N_2,\\
		&\underset{z=\widehat{\xi }_n}{\text{Res}}\,\,W^-\left( z \right) =\left( 0,\ A\left( \widehat{\xi }_n \right) e^{2i\theta \left( \widehat{\xi }_n \right)}\nu _{-,1}\left( x,t,\widehat{\xi }_n \right) \right) ,\ n=1,\cdot \cdot \cdot ,2N_1+2N_2.\ 
	\end{align}
	\subsection{Closing the system}
	\hspace{0.7cm}In order to solve the RH problem, we need to subtract the asymptotic behaviors and the singularities of $W$ at the poles. Thus, jump condition (\ref{e20}) reduces to
	\begin{equation}
		\begin{aligned}
			&W^--e^{im_-\sigma _{\ast}}-\frac{i}{z}\sigma \eta e^{im_-\sigma _{\ast}}\sigma _3Q_{-}^{\ast}-\sum_{n=1}^{2N_1+2N_2}{\frac{\underset{z=\widehat{\xi }_n}{\text{Res}}\,\,W^-}{z-\widehat{\xi }_n}}-\sum_{n=1}^{2N_1+2N_2}{\frac{\underset{z=\xi _n}{\text{Res}}\,\,W^+}{z-\xi _n}} \\
			=\ &W^+-e^{im_-\sigma _{\ast}}-\frac{i}{z}\sigma \eta e^{im_-\sigma _{\ast}}\sigma _3Q_{-}^{\ast}-\sum_{n=1}^{2N_1+2N_2}{\frac{\underset{z=\widehat{\xi }_n}{\text{Res}}\,\,W^-}{z-\widehat{\xi }_n}}-\sum_{n=1}^{2N_1+2N_2}{\frac{\underset{z=\xi _n}{\text{Res}}\,\,W^+}{z-\xi _n}}-W^+J,
		\end{aligned}
	\end{equation}
	according to the Plemelj formula, we have
	\begin{equation}
		\begin{aligned}
			W\left( x,t,z \right) &=e^{im_-\sigma _{\ast}}+\frac{i}{z}\sigma \eta e^{im_-\sigma _{\ast}}\sigma _3Q_{-}^{\ast}+\sum_{n=1}^{2N_1+2N_2}{\frac{\underset{z=\widehat{\xi }_n}{\text{Res}}\,\,W^-}{z-\widehat{\xi }_n}}
			+\sum_{n=1}^{2N_1+2N_2}{\frac{\underset{z=\xi _n}{\text{Res}}\,\,W^+}{z-\xi _n}}\\
			&+\frac{1}{2\pi i}\int_{\varSigma}{\frac{W^+\left( s \right) J\left( s \right)}{s-z}}ds.
		\end{aligned}
	\end{equation}
	When $z\rightarrow\infty $, the asymptotic expansion can be obtained:
	\begin{equation}
		\begin{aligned}\label{e19}
			W\left( x,t,z \right) &=e^{im_-\sigma _{\ast}}+\frac{1}{z}\left\{i \sigma \eta e^{im_-\sigma _{\ast}}\sigma _3Q_{-}^{\ast}+\sum_{n=1}^{2N_1+2N_2}{\left( \underset{\xi _n}{\text{Res}}W^++\underset{\widehat{\xi }_n}{\text{Res}}W^- \right)} \right\}   \\
			&-\frac{1}{z}\left\{ \frac{1}{2\pi i}\int_{\varSigma}{W^+\left( s \right) J\left( s \right)}ds \right\} +O\left( z^{-2} \right) , 
		\end{aligned}
	\end{equation}
	by comparing the formula (\ref{e18}) with the 11 position elements of formula (\ref{e19}), the reconstruction formula is
	\begin{equation}\label{e24}
		q=e^{2im_-}q_-+ie^{im_-}\sum_{n=1}^{2N_1+2N_2}{\frac{a\left( \xi _n \right)}{u_{11}^{'}\left( \xi _n \right)}e^{-2i\theta \left( \xi _n \right)}\nu _{-,12}\left(\xi _n \right) +\frac{e^{im_-}}{2\pi}\int_{\varSigma}{\left( W^+\left( s \right) J\left( s \right) \right) _{11}}ds}.
	\end{equation}
	\subsection{Trace Formulae and $\theta$ Condition}
	\hspace{0.7cm}Since $u_{11}\left( z \right) $ and $u_{22}\left( z \right) $ are analytic in $ D^+$ and $ D^-$, respectively, and the discrete spectrum $\xi _n$ and $\widehat{\xi }_n$ are the simple zeros of $u_{11}\left( z \right) $ and $u_{22}\left( z \right) $, respectively, then, one can find the trace formulae
	\begin{equation}
		u_{11}\left( z \right) =\prod_{n=1}^{2N_1+2N_2}{\frac{z-\xi _n}{z-\widehat{\xi }_n}}e^{i\bar{m}}\exp \left[ -\frac{1}{2\pi i}\int_{\varSigma}{\frac{\log \left[ 1-\rho \left( s \right) \tilde{\rho}\left( s \right) \right]}{s-z}}ds \right],\ z\in D^+ ,
	\end{equation}
	\begin{equation}
		u_{22}\left( z \right) =\prod_{n=1}^{2N_1+2N_2}{\frac{z-\widehat{\xi }_n}{z-\xi _n}}e^{-i\bar{m}}\exp \left[ \frac{1}{2\pi i}\int_{\varSigma}{\frac{\log \left[ 1-\rho \left( s \right) \tilde{\rho}\left( s \right) \right]}{s-z}}ds \right] ,\ z\in D^- .
	\end{equation}
	We consider the case of reflectionless potentials, i.e., $u_{12}\left( z \right) =u_{21}\left( z \right) =0$, the above trace formulae are simplified as
	\begin{equation}
		u_{11}\left( z \right) =\prod_{n=1}^{2N_1+2N_2}{\frac{z-\xi _n}{z-\widehat{\xi }_n}}e^{i\bar{m}},\ z\in D^+ ,\ \ 
		u_{22}\left( z \right) =\prod_{n=1}^{2N_1+2N_2}{\frac{z-\widehat{\xi }_n}{z-\xi _n}}e^{-i\bar{m}},\ z\in D^- .
	\end{equation}
	When $z\rightarrow 0$, the $\theta $ condition can be obtained:
	\begin{equation}
		\frac{q_+}{q_-}=\prod_{n=1}^{2N_1+2N_2}{\frac{\xi _n}{\widehat{\xi }_n}}e^{2i\bar{m}}\exp\left[ -\frac{1}{2\pi i}\int_{\varSigma}{\frac{\log \left[ 1-\rho \left( s \right) \tilde{\rho}\left( s \right) \right]}{s}}ds \right] .
	\end{equation}	
	In the case of reflectionless potentials, it can be simplified as
	\begin{equation}\label{e26}
		\frac{q_+}{q_-}=\prod_{n=1}^{2N_1+2N_2}{\frac{\xi _n}{\widehat{\xi }_n}}e^{2i\bar{m}} .
	\end{equation}
	\subsection{Soliton solutions}
	\hspace{0.7cm}Under the condition of reflectionless potential, the first column of $W^-$ is given by
	\begin{equation}
		\nu _{-,1}\left( \widehat{\xi }_n \right) =\left( \begin{array}{c}
			-\frac{i}{\widehat{\xi }_n}e^{im_-}q_-\\
			e^{-im_-}\\
		\end{array} \right) +\sum_{n=1}^{2N_1+2N_2}{\frac{A\left( \xi _n \right)}{\widehat{\xi }_n-\xi _n}e^{-2i\theta \left( \xi _n \right)}}\nu _{-,2}\left( \xi _n \right) ,
	\end{equation}
	taking the element at position 11, one has
	\begin{equation}\label{e22}
		\nu _{-,11}\left( \widehat{\xi }_n \right) =-\frac{i}{\widehat{\xi }_n}e^{im_-}q_-+\sum_{n=1}^{2N_1+2N_2}{\frac{A\left( \xi _n \right)}{\widehat{\xi }_n-\xi _n}e^{-2i\theta \left( \xi _n \right)}}\nu _{-,12}\left( \xi _n \right) .
	\end{equation}
	Then, by means of the symmetry (\ref{e21}), we derive
	\begin{equation}\label{e23}
		\nu _{-,11}\left( \widehat{\xi }_n \right) =-\frac{i\xi _n}{q_-^{\ast}}\sigma \eta \nu _{-,12}\left( \xi _n \right) .
	\end{equation}
	Comparing Eq.~(\ref{e22}) with Eq.~(\ref{e23}) yields
	$$
	-\frac{i\xi _k}{q_-^{\ast}}\sigma \eta \nu _{-,12}\left( \xi _k \right) =-\frac{i}{\widehat{\xi }_k}e^{im_-}q_-+\sum_{n=1}^{2N_1+2N_2}{\frac{A\left( \xi _n \right)}{\widehat{\xi }_k-\xi _n}e^{-2i\theta \left( \xi _n \right)}}\nu _{-,12}\left( \xi _n \right) ,
	$$
	i.e.,
	$$
	\sum_{n=1}^{2N_1+2N_2}{\left( \frac{A\left( \xi _n \right)}{\widehat{\xi }_k-\xi _n}e^{-2i\theta \left( \xi _n \right)}+\frac{i\xi _k}{q_-^{\ast}}\sigma \eta \delta _{k,n} \right)}\nu _{-,12}\left( \xi _n \right) =\frac{iq_-}{\widehat{\xi }_k}e^{im_-},\ k=1,\cdot \cdot \cdot ,\ 2N_1+2N_2 .
	$$
	By using Cramer's rule we can solve for 
	\begin{equation}\label{e25}
		\nu _{-,12}\left( \xi _n \right) =\frac{\det \left( G_1,\cdot \cdot \cdot ,G_{n-1},B,G_{n+1},\cdot \cdot \cdot ,G_{2N_1+2N_2} \right)}{\det G} , 
	\end{equation}
	where
	\begin{equation}
		X=\left( \nu _{-,12}\left( \xi _1 \right) ,\cdot \cdot \cdot ,\nu _{-,12}\left( \xi _{2N_1+2N_2} \right) \right) ^T,\ B=\left( \frac{iq_-}{\widehat{\xi }_1}e^{im_-},\cdot \cdot \cdot ,\frac{iq_-}{\ \ \ \widehat{\xi }_{2N_1+2N_2}}e^{im_-} \right) ^T ,
	\end{equation}
	\begin{equation}
		G=g_{kn}=\frac{A\left( \xi _n \right)}{\widehat{\xi }_k-\xi _n}e^{-2i\theta \left( \xi _n \right)}+\frac{i\xi _k}{q_-^{\ast}}\sigma \eta \delta _{k,n}\ .
	\end{equation}
	From Eqs.~(\ref{e24}) and (\ref{e25}), the solution of Eq.~(\ref{e1}) under nonzero boundary conditions is given by
	\begin{equation}
		\begin{aligned}
			q&=e^{2im_-}q_-+ie^{im_-}\sum_{n=1}^{2N_1+2N_2}{\frac{a\left( \xi _n \right)}{u_{11}^{'}\left( \xi _n \right)}e^{-2i\theta \left( \xi _n \right)}\nu _{-,12}\left( x,t,\xi _n \right)} \\
			&=e^{2im_-}q_--ie^{im_-}\frac{\det G^{aug}}{\det G} , 
		\end{aligned}
	\end{equation}
	where
	$$
	\det G^{aug}=\left| \begin{matrix}
		0&		Y\\
		B&		G\\
	\end{matrix} \right|,\ Y_n=\frac{a\left( \xi _n \right)}{u_{11}^{'}\left( \xi _n \right)}e^{-2i\theta \left( \xi _n \right)} .
	$$ 
	
	Since $\sigma$ can be unified through a transformation, therefore we will exclusively focus on the case where $\sigma=-1$. 
	
Case 1: $\sigma=-1,\ \eta=-1$. 
In this case, $\theta$ condition (\ref{e26}) reduces to 
\begin{equation}
	-\frac{q_-^{\ast}}{q_-}=\prod_{n=1}^{N_1}{\frac{\zeta _n^4}{q_0^4}}e^{4i\sum_{k=1}^{N_2}{arg \delta _k}}e^{2i\bar{m}}.
\end{equation}
$\bullet$ For $N_1=0,\ N_2=2,$ we take $q_-=1,\ \delta_1=e^{i\theta_{1}},\ \delta_2=e^{i\theta_{2}},\ \theta_{1},\theta_{2} \in \left( 0,\frac{\pi}{2} \right).$ Then, from Eq.~(\ref{e27}), we know that $a^2\left( \pm\delta _1 \right) =a^2\left( \pm\delta _2 \right) =1,\ a\left( \delta _1 \right) a\left( -\delta _1 \right) =a\left( \delta _2 \right) a\left( -\delta _2 \right) =-1$. Select the specific value and we get the dark-dark soliton solution as shown in Fig. 2.
\begin{figure}[H]
	\centering
	\begin{minipage}[c]{0.31\textwidth} 
		\centering
		\includegraphics[height=4cm,width=\textwidth]{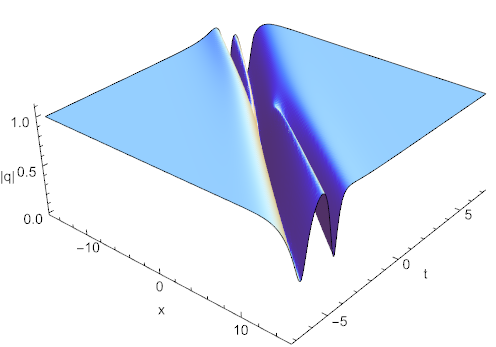} 
		\centerline{(a)}
	\end{minipage}
	\begin{minipage}[c]{0.31\textwidth}
		\centering
		\includegraphics[height=4cm,width=\textwidth]{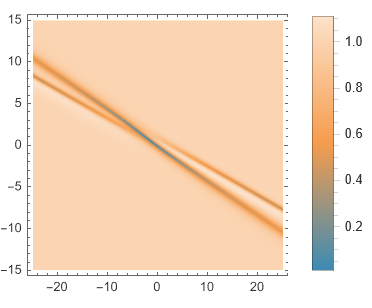}
		\centerline{(b)}
	\end{minipage}
	\begin{minipage}[c]{0.31\textwidth}
		\centering
		\includegraphics[height=4cm,width=\textwidth]{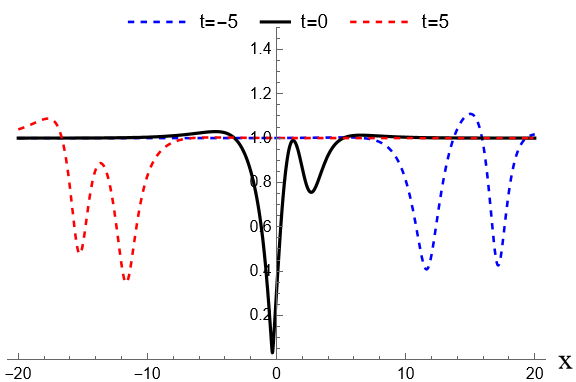}
		\centerline{(c)}
	\end{minipage}
	\caption{Dark-dark soliton solution with $q_-=1,\ \delta_1=e^{\frac{i\pi}{6}},\ \delta_2=e^{\frac{i\pi}{12}}.\ a\left( \delta _1 \right)=a\left( \delta _2 \right)=1,\ a\left( -\delta _1 \right)=a\left( -\delta _2 \right)=-1.$ (a) the three-dimensional plot, (b) the two-dimensional density plot, (c) plots of $q(x)$ with three values of t.}
\end{figure}
$\bullet$ For $N_1=2,\ N_2=0,$ we let $q_-=1,\,\,\zeta_{1}=|\zeta_{1}|e^{i\theta _1},\,\,\zeta_{2}=|\zeta_{2}|e^{i\theta _2},\ \theta _1,\theta _2\in \left( 0,\frac{\pi}{2} \right) ,\ |\zeta_{1}||\zeta_{2}|=q_0^2.\ $Then, from Eq.~(\ref{e27}), we know that $a^2\left( \pm\zeta_{1} \right) =a^2\left( \pm\zeta_{2} \right) =1,\ a\left( \zeta_{1} \right) a\left( -\zeta_{1} \right) = a\left( \zeta_{2} \right) a\left( -\zeta_{2} \right) =-1$. Select the specific value and we get the breather solution as shown in Fig. 3.
\begin{figure}[H]
	\centering
	\begin{minipage}[c]{0.31\textwidth} 
		\centering
		\includegraphics[height=4cm,width=\textwidth]{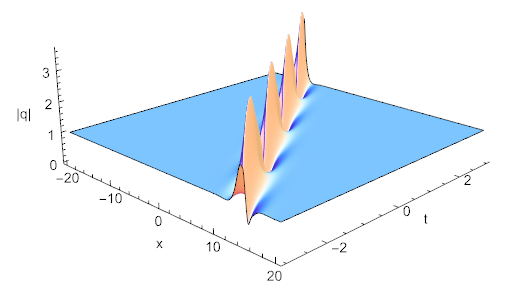} 
		\centerline{(a)}
	\end{minipage}
	\begin{minipage}[c]{0.31\textwidth}
		\centering
		\includegraphics[height=4cm,width=\textwidth]{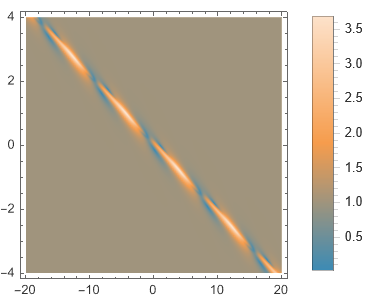}
		\centerline{(b)}
	\end{minipage}
	\begin{minipage}[c]{0.31\textwidth}
		\centering
		\includegraphics[height=4cm,width=\textwidth]{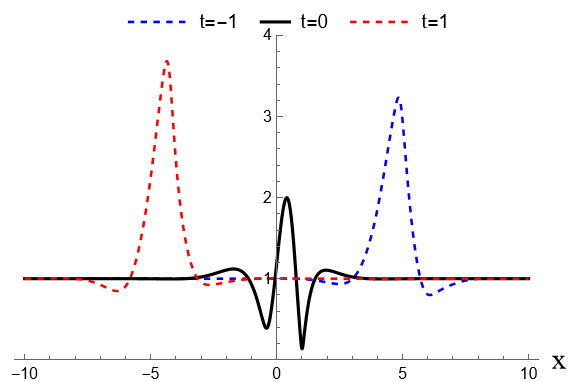}
		\centerline{(c)}
	\end{minipage}
	\caption{Breather solution with $q_-=1,\ \zeta_{1}=2e^{\frac{i\pi}{8}},\ \zeta_{2}=\frac{1}{2}e^{\frac{i\pi}{8}}.\ a\left( \zeta_{1} \right)=a\left( \zeta_{2} \right)=1,\ a\left( -\zeta_{1} \right)=a\left( -\zeta_{2} \right)=-1.$ (a) the three-dimensional plot, (b) the two-dimensional density plot, (c) plots of $q(x)$ with three values of t.}
\end{figure}

Case 2: $\sigma=-1 , \eta=1$. 
In this case, $\theta$ condition (\ref{e26}) reduces to 
\begin{equation}
	\frac{q_-^{\ast}}{q_-}=\prod_{n=1}^{N_1}{\frac{\zeta_n^4}{q_0^4}}e^{4i\sum_{k=1}^{N_2}{arg\delta _k}}e^{2i\bar{m}}.
\end{equation}

$\bullet$ For $N_1=2,\ N_2=0,$ we let $q_-=1,\,\,\zeta_{1}=|\zeta_{1}|e^{i\theta _1},\,\,\zeta_{2}=|\zeta_{2}|e^{i\theta _2},\ \theta _1,\theta _2\in \left( 0,\frac{\pi}{2} \right) ,\ |\zeta_{1}||\zeta_{2}|=q_0^2.\ $Then, from Eq.~(\ref{e27}), we know that $a^2\left( \pm\zeta_{1} \right) =a^2\left( \pm\zeta_{2} \right) =1,\ a\left( \zeta_{1} \right) a\left( -\zeta_{1} \right) = a\left( \zeta_{2} \right) a\left( -\zeta_{2} \right) =-1$. Select the specific value and we get the breather solution as shown in Fig. 4.
\begin{figure}[H]
	\centering
	\begin{minipage}[c]{0.31\textwidth} 
		\centering
		\includegraphics[height=4cm,width=\textwidth]{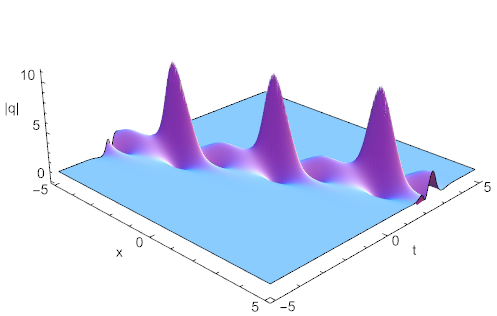} 
		\centerline{(a)}
	\end{minipage}
	\begin{minipage}[c]{0.31\textwidth}
		\centering
		\includegraphics[height=4cm,width=\textwidth]{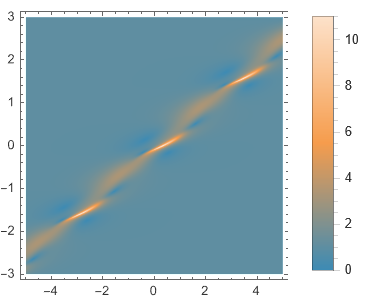}
		\centerline{(b)}
	\end{minipage}
	\begin{minipage}[c]{0.31\textwidth}
		\centering
		\includegraphics[height=4cm,width=\textwidth]{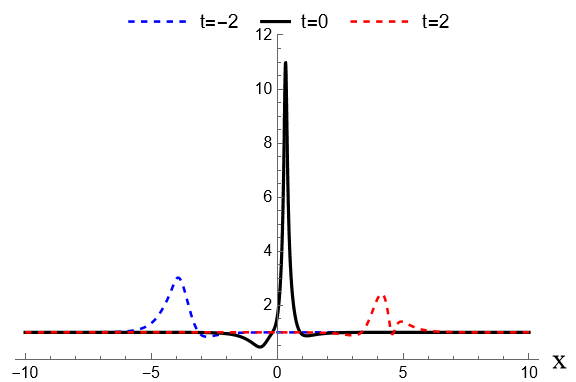}
		\centerline{(c)}
	\end{minipage}
	\caption{Breather solution with $q_-=1,\ \zeta_1=2e^{\frac{i\pi}{4}},\ \zeta_2=\frac{1}{2}e^{\frac{i\pi}{4}},\ a\left( \zeta_{1} \right)=a\left( \zeta_{2} \right)=1,\ a\left( -\zeta_{1} \right)=a\left( -\zeta_{2} \right)=-1.$ (a) the three-dimensional plot, (b) the two-dimensional density plot, (c) plots of $q(x)$ with three values of t.}
\end{figure}
$\bullet$ As $N_1=0,\ N_2=2,$ we let $q_-=1,\ \delta_1=e^{i\theta_{1}},\ \delta_2=e^{i\theta_{2}},\ \theta_{1},\theta_{2} \in \left( 0,\frac{\pi}{2} \right).$ Then, from formula (\ref{e27}), we know that $a^2\left( \pm\delta _1 \right) =a^2\left( \pm\delta _2 \right) =1,\ a\left( \delta _1 \right) a\left( -\delta _1 \right) =a\left( \delta _2 \right) a\left( -\delta _2 \right) =-1$. By selecting different values, we get the bright-dark soliton solution shown in Fig. 5.
\begin{figure}[H]
	\centering
	\begin{minipage}[c]{0.31\textwidth} 
		\centering
		\includegraphics[height=4cm,width=\textwidth]{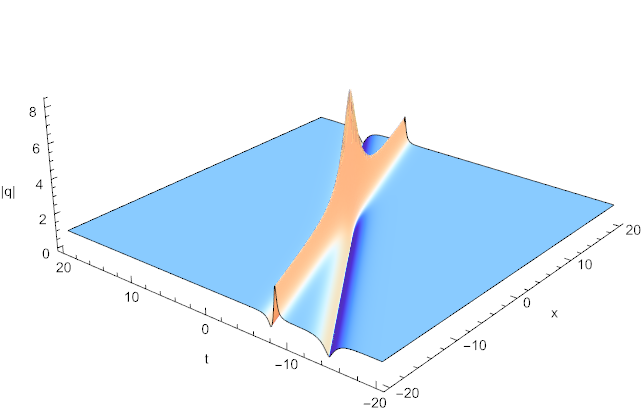} 
		\centerline{(a)}
	\end{minipage}
	\begin{minipage}[c]{0.31\textwidth}
		\centering
		\includegraphics[height=4cm,width=\textwidth]{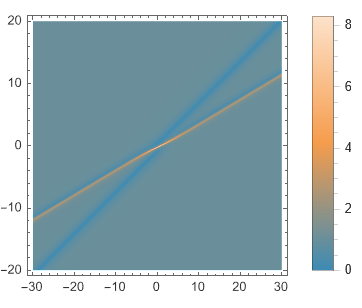}
		\centerline{(b)}
	\end{minipage}
	\begin{minipage}[c]{0.31\textwidth}
		\centering
		\includegraphics[height=4cm,width=\textwidth]{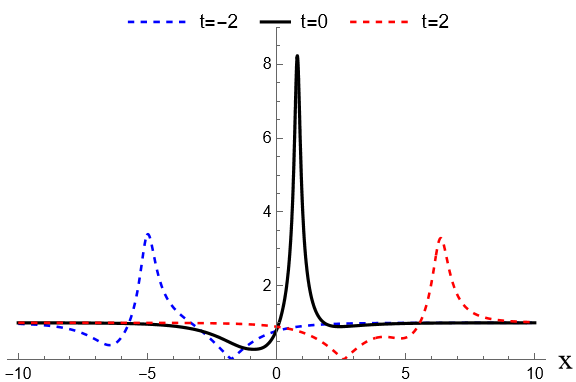}
		\centerline{(c)}
	\end{minipage}
	\caption{Bright-dark soliton solution with $q_-=1,\ \delta_1=e^{\frac{i\pi}{6}},\ \delta_2=e^{\frac{i\pi}{3}}.\ a\left( \delta _1 \right)=a\left( \delta _2 \right)=1,\ a\left( -\delta _1 \right)=a\left( -\delta _2 \right)=-1.$ (a) the three-dimensional plot, (b) the two-dimensional density plot, (c) plots of $q(x)$ with three values of t.}
\end{figure}
	
	\section{Inverse scattering problem with double-pole }
	\subsection{Discrete spectrum and residue conditions}
	\hspace{0.7cm}We assume that $u_{11} (z) $ has $N_1$ double zeros in $D _{+}\bigcap\left\{ z\in \mathbb{C}\mid \text{Re}z>0,\text{Im}z>0,|z| \ne q_0 \right\} $, expressed as $\zeta_{n},n=1,2,\cdot\cdot\cdot,N_1$. Then we suppose that $u_{11} (z) $ has $N_2$ double zeros in $D _{+}\bigcap \left\{ z\in \mathbb{C}\mid \text{Re}z>0,\text{Im}z>0,|z| =q_0 \right\} $, expressed as $\delta_{n} , n=1,2,\cdot\cdot\cdot,N_2$. Similar to Section 3, we have the discrete spectrum $Z=\left\{ \xi _n,\ \widehat{\xi }_n,\ n=1,\cdot \cdot \cdot ,\ 2N_1+ 2N_2 \right\} $, 
	and if $u_{11}\left( \xi _n \right) =0$, then $u_{11}^{'}\left( \xi _n \right) =u_{22}\left( \hat{\xi}_n \right) =\,u_{22}^{'}\left( \hat{\xi}_n \right) =0$. It follows from the Wronskian determinants (\ref{e16}-\ref{e17}) that the following equations can be obtained:
	\begin{subequations}
		\begin{align}
			&\varPsi _{+,1}\left( z \right) =b\left( z \right) \varPsi _{-,2}\left( z \right) ,\,\,z\in Z\cap D_+, \\
			&\varPsi _{+,2}\left( z \right) =b\left( z \right) \varPsi _{-,1}\left( z \right) ,\ z\in Z\cap D_-, \\
			&\varPsi _{+,1}^{'}\left( z \right) =b\left( z \right) \varPsi _{-,2}^{'}\left( z \right) +d\left( z \right) \varPsi _{-,2}\left( z \right) ,\ z\in Z\cap D_+, \\
			& \varPsi _{+,2}^{'}\left( z \right) =b\left( z \right) \varPsi _{-,1}^{'}\left( z \right) +d\left( z \right) \varPsi _{-,1}\left( z \right) ,\ z\in Z\cap D_-,		
		\end{align}
	\end{subequations}
	where $b\left( z \right)$ and $d\left( z \right)$ are constants independent of $x$ and $t$.
	
	In this case, according to the matrix function $W$ and Eq.~(\ref{e35}), the residue of $W$ and the coefficient of the negative quadratic power in the Laurent expansion of $W$ jointly affect the solution of the RH problem. Through calculation, their expressions can be obtained as follows:
	\begin{subequations}
		\begin{align}\label{e29}
			&\underset{z=\xi _n}{\text{Res}}\,\,W^+\left( z \right) =\left( A\left( \xi _n \right) e^{-2i\theta \left( \xi _n \right)}\left[ \nu _{-,2}^{'}\left( \xi _n \right) +\left( B\left( \xi _n \right) -2i\theta ^{'}\left( \xi _n \right) \right) \nu _{-,2}\left( \xi _n \right) \right] ,\,\,0 \right) ,\\
			&\underset{z=\hat{\xi}_n}{\text{Res}}\,\,W^-\left( z \right)=\left( 0,A\left( \hat{\xi}_n \right) e^{2i\theta \left( \hat{\xi}_n \right)}\left[ \nu _{-,1}^{'}\left( \hat{\xi}_n \right) +\left( B\left( \hat{\xi}_n \right) +2i\theta ^{'}\left( \hat{\xi}_n \right) \right) \nu _{-,1}\left( \hat{\xi}_n \right) \right] \right) , \\
			&\underset{z=\xi _n}{L_{-,2}}\,\,\,W^+\left( z \right)=\left( A\left( \xi _n \right) e^{-2i\theta \left( \xi _n \right)}\nu _{-,2}\left( \xi _n \right) ,0 \right) , \\
			&\underset{z=\hat{\xi}_n}{L_{-,2}}\,\,\,W^-\left( z \right)=\left( 0,A\left( \hat{\xi}_n \right) e^{2i\theta \left( \hat{\xi}_n \right)}\nu _{-,1}\left( \hat{\xi}_n \right) \right) ,
		\end{align}
	\end{subequations}	
	where the prime represents the partial derivative with respect to $z$, and
	\begin{align}
		&A\left( \xi _n \right) =\frac{2b\left( \xi _n \right)}{u_{11}^{''}\left( \xi _n \right)},\ A\left( \hat{\xi}_n \right) =\frac{2b\left( \hat{\xi}_n \right)}{u_{22}^{''}\left( \hat{\xi}_n \right)}, \\
		&B\left( \xi _n \right) =\frac{d\left( \xi _n \right)}{b\left( \xi _n \right)}-\frac{u_{11}^{'''}\left( \xi _n \right)}{3u_{11}^{''}\left( \xi _n \right)},\,\,B\left( \hat{\xi}_n \right) =\frac{d\left( \hat{\xi}_n \right)}{b\left( \hat{\xi}_n \right)}-\frac{u_{22}^{'''}\left( \hat{\xi}_n \right)}{3u_{22}^{''}\left( \hat{\xi}_n \right)}.		
	\end{align}
	
	\subsection{Symmetry relations}
	\hspace{0.7cm}For $z\in Z$, the two symmetry relations for $b\left( z \right)$, $d\left( z \right)$, $u_{11}^{'}\left( z \right)$, $u_{11}^{''}\left( z \right)$, $u_{22}^{'}\left( z \right)$ and $u_{22}^{''}\left( z \right)$ are as follows: 
	
	$\bullet$ The first symmetry relation:
	\begin{subequations}\label{e40}
			\begin{align}
	    	&b\left( z \right) =\frac{\sigma}{b\left( -z \right)},\ d\left( z \right) =\sigma b^{-2}\left( -z \right) d\left( -z \right) ,\\
			&u_{11}^{'}\left( z \right) =-u_{11}^{'}\left( -z \right) ,\ u_{22}^{'}\left( z \right) =-u_{22}^{'}\left( -z \right) ,\\
			&u_{11}^{''}\left( z \right) =u_{11}^{''}\left( -z \right) ,\ u_{22}^{''}\left( z \right) =u_{22}^{''}\left( -z \right) .			
		\end{align}
	\end{subequations}	
	
	$\bullet$ The second symmetry relation:
	\begin{subequations}
		\begin{align}
			&b\left( z \right) =-\frac{\sigma}{b\left( \frac{\sigma \eta q_{0}^{2}}{z} \right)},\ d\left( z \right) =\frac{\eta q_{0}^{2}}{z^2}d\left( \frac{\sigma \eta q_{0}^{2}}{z} \right), \\
			&u_{11}^{'}\left( z \right) =-\frac{\sigma q_{+}^{2}}{z^2}u_{22}^{'}\left( \frac{\sigma \eta q_{0}^{2}}{z} \right) ,\ u_{11}^{''}\left( z \right) =\frac{2\sigma q_{+}^{2}}{z^3}u_{22}^{'}\left( \frac{\sigma \eta q_{0}^{2}}{z} \right) +\frac{\eta q_{+}^{2}q_{0}^{2}}{z^4}u_{22}^{''}\left( \frac{\sigma \eta q_{0}^{2}}{z} \right) ,\\			
			&u_{22}^{'}\left( z \right) =-\frac{\sigma q_{-}^{2}}{z^2}u_{11}^{'}\left( \frac{\sigma \eta q_{0}^{2}}{z} \right) ,\ u_{22}^{''}\left( z \right) =\frac{2\sigma q_{-}^{2}}{z^3}u_{11}^{'}\left( \frac{\sigma \eta q_{0}^{2}}{z} \right) +\frac{\eta q_{-}^{2}q_{0}^{2}}{z^4}u_{11}^{''}\left( \frac{\sigma \eta q_{0}^{2}}{z} \right) .
		\end{align}
	\end{subequations}	
	\subsection{Closing the system}
	\hspace{0.7cm}Similarly, to solve the RH problem, we must remove the asymptotic values and the singularity for $W$ at the poles,
	\begin{align*}
		&W^--e^{im_-\sigma _{\ast}}\left( I+\frac{i}{z}\sigma \eta \sigma _3Q_{-}^{\ast} \right)-\sum_{n=1}^{2N_1+2N_2}{\left[ \frac{\underset{z=\widehat{\xi }_n}{\text{Res}}\,\,W^-}{z-\widehat{\xi }_n}+\frac{\underset{z=\xi _n}{\text{Res}}\,\,W^+}{z-\xi _n}+\frac{\underset{z=\hat{\xi}_n}{L_{-,2}}\,\,\,W^-}{\left( z-\widehat{\xi }_n \right) ^2}+\frac{\underset{z=\xi _n}{L_{-,2}}\,\,W^+}{\left( z-\xi _n \right) ^2} \right]} \\
		=\ &W^+-e^{im_-\sigma _{\ast}}\left( I+\frac{i}{z}\sigma \eta \sigma _3Q_{-}^{\ast} \right)-\sum_{n=1}^{2N_1+2N_2}{\left[ \frac{\underset{z=\widehat{\xi }_n}{\text{Res}}\,\,W^-}{z-\widehat{\xi }_n}+\frac{\underset{z=\xi _n}{\text{Res}}\,\,W^+}{z-\xi _n}+\frac{\underset{z=\hat{\xi}_n}{L_{-,2}}\,\,\,W^-}{\left( z-\widehat{\xi }_n \right) ^2}+\frac{\underset{z=\xi _n}{L_{-,2}}\,\,W^+}{\left( z-\xi _n \right) ^2} \right]}-W^+J,
	\end{align*}
	by the Plemelj formula, we have
	\begin{equation}\label{e30}
		\begin{aligned}
			W\left( x,t,z \right) =&e^{im_-\sigma _{\ast}}+\frac{i}{z}\sigma \eta e^{im_-\sigma _{\ast}}\sigma _3Q_{-}^{\ast}+\sum_{n=1}^{2N_1+2N_2}{\frac{\underset{z=\widehat{\xi }_n}{\text{Res}}\,\,W^-}{z-\widehat{\xi }_n}}+\sum_{n=1}^{2N_1+2N_2}{\frac{\underset{z=\xi _n}{\text{Res}}\,\,W^+}{z-\xi _n}} \\
			+&\sum_{n=1}^{2N_1+2N_2}{\frac{\underset{z=\hat{\xi}_n}{L_{-,2}}\,\,\,W^-}{\left( z-\widehat{\xi }_n \right) ^2}}+\sum_{n=1}^{2N_1+2N_2}{\frac{\underset{z=\xi _n}{L_{-,2}}\,\,W^+}{\left( z-\xi _n \right) ^2}}+\frac{1}{2\pi i}\int_{\varSigma}{\frac{W^+\left( s \right) J\left( s \right)}{s-z}}ds.	
		\end{aligned}
	\end{equation}
	Substituting Eq.~(\ref{e29}) into Eq.~(\ref{e30}), we find the following equation:
	\begin{equation}\label{e28}
		\begin{aligned}
			W\left( x,t,z \right) =&e^{im_-\sigma _{\ast}}\left( I+\frac{i}{z}\sigma \eta \sigma _3Q_{-}^{\ast} \right) +\frac{1}{2\pi i}\int_{\varSigma}{\frac{W^+\left( s \right) J\left( s \right)}{s-z}}ds\\
			+&\sum_{n=1}^{2N_1+2N_2}{\left( C_n\left( z \right) \left[ \nu _{-,2}^{'}\left( \xi _n \right) +\left( D_n+\frac{1}{z-\xi _n} \right) \nu _{-,2}\left( \xi _n \right) \right],0 \right)} \\
			+&\sum_{n=1}^{2N_1+2N_2}{\left( 0,\hat{C}_n\left( z \right) \left[ \nu _{-,1}^{'}\left( \hat{\xi}_n \right) +\left( \hat{D}_n+\frac{1}{z-\hat{\xi}_n} \right) \nu _{-,1}\left( \hat{\xi}_n \right) \right] \right)},		
		\end{aligned}
	\end{equation}
	where
	\begin{align}
		&C_n\left( z \right) =\frac{A\left( \xi _n \right)}{z-\xi _n}e^{-2i\theta \left( \xi _n \right)},\ D_n=B\left( \xi _n \right) -2i\theta ^{'}\left( \xi _n \right), \\
		&\hat{C}_n\left( z \right) =\frac{A\left( \hat{\xi}_n \right)}{z-\hat{\xi}_n}e^{2i\hat{\theta}\left( \hat{\xi}_n \right)},\ \hat{D}_n=B\left( \hat{\xi}_n \right) +2i\theta ^{'}\left( \hat{\xi}_n \right).		
	\end{align}
	By comparing the formula (\ref{e18}) with the 11 position elements of the formula (\ref{e28}), the reconstruction formula is	
	\begin{equation}\label{e34}
		\begin{aligned}
			q\left( x,t \right) =&ie^{im_-}\sum_{n=1}^{2N_1+2N_2}{\left[ A\left( \xi _n \right) e^{-2i\theta \left( \xi _n \right)}\left( \nu _{-,12}^{'}\left( \xi _n \right) +D_n\nu _{-,12}\left( \xi _n \right) \right) \right]}\\
			+&e^{2im_-}q_--\frac{e^{im_-}}{2\pi}\int_{\varSigma}{\left( W^+\left( s \right) J\left( s \right) \right) _{11}}ds.		
		\end{aligned}
	\end{equation}

	\subsection{Trace Formulae and $\theta$ Condition}
	\hspace{0.7cm}We have the following trace formulae:
	\begin{equation}
		u_{11}\left( z \right) =\prod_{n=1}^{2N_1+2N_2}{\left( \frac{z-\xi _n}{z-\widehat{\xi }_n} \right) ^2}e^{i\bar{m}}\exp \left[ -\frac{1}{2\pi i}\int_{\varSigma}{\frac{\log \left[ 1-\rho \left( s \right) \tilde{\rho}\left( s \right) \right]}{s-z}}ds \right] ,\,\,z\in D^+,
	\end{equation}
	
	\begin{equation}
		u_{22}\left( z \right) =\prod_{n=1}^{2N_1+2N_2}{\left( \frac{z-\widehat{\xi }_n}{z-\xi _n} \right) ^2}e^{-i\bar{m}}\exp \left[ \frac{1}{2\pi i}\int_{\varSigma}{\frac{\log \left[ 1-\rho \left( s \right) \tilde{\rho}\left( s \right) \right]}{s-z}}ds \right] ,\,\,z\in D^-.
	\end{equation}
	
	When $z\rightarrow 0$, the $\theta $ condition can be obtained:
	\begin{equation}\label{e39}
		\frac{q_+}{q_-}=\prod_{n=1}^{2N_1+2N_2}{\frac{\xi _{n}^{2}}{\hat{\xi}_{n}^{2}}}e^{2i\bar{m}}\exp\left[ -\frac{1}{2\pi i}\int_{\varSigma}{\frac{\log \left[ 1-\rho \left( s \right) \tilde{\rho}\left( s \right) \right]}{s}}ds \right].
       	\end{equation}		
	
	\subsection{Soliton solutions}	
	\hspace{0.7cm} Due to the symmetry (\ref{e21}), $\nu _{-,1}\left( \hat{\xi}_n \right)$ and $\nu _{-,1}^{'}\left( \hat{\xi}_n \right)$ are determined by $\nu _{-,2}\left( {\xi}_n \right)$ and $\nu _{-,2}^{'}\left( {\xi}_n \right)$ as 
	\begin{equation}\label{e31}
		\nu _{-,1}\left( \hat{\xi}_n \right) =\frac{-i\sigma \xi _n}{q_+}\nu _{-,2}\left( \xi _n \right) ,\ \nu _{-,1}^{'}\left( \hat{\xi}_n \right) =\frac{i\xi _{n}^{2}}{q_{-}^{*}q_{0}^{2}}\nu _{-,2}\left( \xi _n \right) +\frac{i\xi _{n}^{3}}{q_{-}^{*}q_{0}^{2}}\nu _{-,2}^{'}\left( \xi _n \right) .
	\end{equation}
	Taking the first column of $W^{-}$ and combining Eq.~(\ref{e31}), we have
	\begin{equation}\label{e32}
		\begin{aligned}
			&\sum_{n=1}^{2N_1+2N_2}{\left[ C_n\left( \hat{\xi}_k \right) \nu _{-,2}^{'}\left( \xi _n \right) +\left[ C_n\left( \hat{\xi}_k \right) \left( D_n+\frac{1}{\hat{\xi}_k-\xi _n} \right) +\frac{i\sigma \xi _k}{q_+}\delta _{k,n} \right] \nu _{-,2}\left( \xi _n \right) \right]} \\
			=&-e^{im_-\sigma _*}\left( \begin{array}{c}
				1\\
				\frac{-iq_-}{\hat{\xi}_k}\\
			\end{array} \right) -\frac{1}{2\pi i}\int_{\varSigma}{\frac{\left( W^+\left( s \right) J\left( s \right) \right) _1}{s-\hat{\xi}_k}}ds,
		\end{aligned}
	\end{equation}
	then, taking the derivative of Eq.~(\ref{e32}), we get
	\begin{equation}\label{e33}
		\begin{aligned}
			&\sum_{n=1}^{2N_1+2N_2}{\left[ \left( \frac{C_n\left( \hat{\xi}_k \right)}{\hat{\xi}_k-\xi _n}+\frac{i\xi _{k}^{3}\delta _{k,n}}{q_{-}^{*}q_{0}^{2}} \right) \nu _{-,2}^{'}\left( \xi _n \right) +\left[ \frac{C_n\left( \hat{\xi}_k \right)}{\hat{\xi}_k-\xi _n}\left( D_n+\frac{2}{\hat{\xi}_k-\xi _n} \right) +\frac{i\xi _{k}^{2}\delta _{k,n}}{q_{-}^{*}q_{0}^{2}} \right] \nu _{-,2}\left( \xi _n \right) \right]} \\
			=&e^{im_-\sigma _*}\left( \begin{array}{c}
				0\\
				\frac{iq_-}{\hat{\xi}_{k}^{2}}\\
			\end{array} \right) +\frac{1}{2\pi i}\int_{\varSigma}{\frac{\left( W^+\left( s \right) J\left( s \right) \right) _1}{s-\hat{\xi}_k}}ds.
		\end{aligned}
	\end{equation}
	
	Under the reflectionless potential, by solving Eqs.~(\ref{e32}) and (\ref{e33}), we ultimately obtained the equations of $\nu _{-,12}\left( {\xi}_n \right)$ and $\nu _{-,12}^{'}\left( {\xi}_n \right)$. Combining with Eq.~(\ref{e34}), the expression for $q(x, t)$ can be written in matrix form:
	\begin{equation}\label{e36}
		q=q_-e^{2im_-}\left( 1+\frac{\det H}{\det R} \right) ,
	\end{equation}
	where
	\begin{align*}
		&H=\left( \begin{matrix}
			0&		\alpha ^T\\
			\beta&		R\\
		\end{matrix} \right) ,\ 
		\beta =\left( \begin{array}{c}
			\beta ^{\left( 1 \right)}\\
			\beta ^{\left( 2 \right)}\\
		\end{array} \right) ,\ 
		\beta ^{\left( 1 \right)}=\left( \frac{1}{\hat{\xi}_j} \right) _{(2N_1+2N_2)\times 1},\ \beta ^{\left( 2 \right)}=\left( \frac{1}{\hat{\xi}_j^2} \right) _{(2N_1+2N_2)\times 1},\,\\
		&\alpha =\left( \alpha ^{\left( 1 \right)},\alpha ^{\left( 2 \right)} \right) ^T,\alpha ^{\left( 1 \right)}=\left( A\left( \xi _n \right) e^{-2i\theta \left( \xi _n \right)}D_n \right) _{1\times (2N_1+2N_2)}\ ,\alpha ^{\left( 2 \right)}=\left( A\left( \xi _n \right) e^{-2i\theta \left( \xi _n \right)} \right) _{1\times (2N_1+2N_2)},\\
		& R=\left[ \begin{matrix}
			R^{\left( 1,1 \right)}&		R^{\left( 1,2 \right)}\\
			R^{\left( 2,1 \right)}&		R^{\left( 2,2 \right)}\\
		\end{matrix} \right] ,\,\,R^{\left( m,s \right)}=\left( R_{k,n}^{\left( m,s \right)} \right) _{(2N_1+2N_2)\times (2N_1+2N_2)},\ m,s=1,2, 
		\end{align*}
		\begin{align*}
		&R_{k,n}^{\left( 1,1 \right)}=C_n\left( \hat{\xi}_k \right) \left( D_n+\frac{1}{\hat{\xi}_k-\xi _n} \right) +\frac{i\sigma \xi _k}{q_+}\delta _{k,n},\ R_{k,n}^{\left( 1,2 \right)}=C_n\left( \hat{\xi}_k \right) , \\
		&	R_{k,n}^{\left( 2,1 \right)}=\frac{C_n\left( \hat{\xi}_k \right)}{\hat{\xi}_k-\xi _n}\left( D_n+\frac{2}{\hat{\xi}_k-\xi _n} \right) +\frac{i\xi _{k}^{2}}{q_{-}^{*}q_{0}^{2}}\delta _{k,n},\ R_{k,n}^{\left( 2,2 \right)}=\frac{C_n\left( \hat{\xi}_k \right)}{\hat{\xi}_k-\xi _n}+\frac{i\xi _{k}^{3}}{q_{-}^{*}q_{0}^{2}}\delta _{k,n}.
	\end{align*}
	Because of the term $m_{_-}=-\frac{1}{2}\int\limits_{-\infty}^x{\left( \sigma \eta q_{0}^{2}-\sigma q\left( y,t \right) q\left( -y,-t \right) \right) dy}$, the expression (\ref{e36}) is an implicit solution. To get the explicit solution, we have
	\begin{equation}
		W=M_-+\lambda\left( z \right) \sum_{n=1}^{2N_1+2N_2}{\left( \frac{\underset{z=\widehat{\xi }_n}{\text{Res}}\left[ \frac{W}{\lambda\left( z \right)} \right]}{z-\hat{\xi}_n}+\frac{\underset{z=\xi _n}{\text{Res}}\left[ \frac{W}{\lambda\left( z \right)} \right]}{z-\xi _n}+\frac{\underset{z=\xi _n}{L_{-,2}}\left[ \frac{W}{\lambda\left( z \right)} \right]}{\left( z-\xi _n \right) ^2}+\frac{\underset{z=\hat{\xi}_n}{L_{-,2}}\left[ \frac{W}{\lambda\left( z \right)} \right] \,}{\left( z-\hat{\xi}_n \right) ^2} \right)}.
	\end{equation}
	Through similar solving methods with Eq.~(\ref{e36}), we yield another implicit formula
	\begin{equation}\label{e37}
		q=q_-e^{im_-}\left( e^{im_-}+\frac{\det \hat{H}}{\det \hat{R}} \right),
	\end{equation}
	where
	\begin{align*}
		& \hat{H}=\left( \begin{matrix}
			0&		\alpha ^T\\
			\beta&		\hat{R}\\
		\end{matrix} \right) ,\ 
		\hat{R}=\left[ \begin{matrix}
			\hat{R}^{\left( 1,1 \right)}&		\hat{R}^{\left( 1,2 \right)}\\
			\hat{R}^{\left( 2,1 \right)}&		\hat{R}^{\left( 2,2 \right)}\\
		\end{matrix} \right] ,\,\,\hat{R}^{\left( m,s \right)}=\left( \hat{R}_{k,n}^{\left( m,s \right)} \right) _{(2N_1+2N_2)\times (2N_1+2N_2)},\ m,s=1,2, \\
		&\hat{R}_{k,n}^{\left( 1,1 \right)}=\frac{C_n\left( \hat{\xi}_k \right) \lambda\left( \hat{\xi}_k \right)}{\lambda\left( \xi _n \right)}\left( D_n+\frac{1}{\hat{\xi}_k-\xi _n}-\frac{\lambda^{'}\left( \xi _n \right)}{\lambda\left( \xi _n \right)} \right) +\frac{i\sigma \xi _k}{q_+}\delta _{k,n},\,\,\hat{R}_{k,n}^{\left( 1,2 \right)}=\frac{C_n\left( \hat{\xi}_k \right) \lambda\left( \hat{\xi}_k \right)}{\lambda\left( \xi _n \right)}, \\
		&\hat{R}_{k,n}^{\left( 2,1 \right)}=\frac{C_n\left( \hat{\xi}_k \right)}{\lambda\left( \xi _n \right)}\left[ \frac{\lambda\left( \hat{\xi}_k \right)}{\hat{\xi}_k-\xi _n}\left( D_n+\frac{2}{\hat{\xi}_k-\xi _n}-\frac{\lambda^{'}\left( \hat{\xi}_k \right)}{\lambda\left( \xi _n \right)} \right) -\lambda^{'}\left( \hat{\xi}_k \right) \left( D_n+\frac{1}{\hat{\xi}_k-\xi _n}-\frac{\lambda^{'}\left( \xi _n \right)}{\lambda\left( \xi _n \right)} \right) \right] +\frac{i\xi _{k}^{2}}{q_{-}^{*}q_{0}^{2}}\delta _{k,n},\\
		&\hat{R}_{k,n}^{\left( 2,2 \right)}=\frac{C_n\left( \hat{\xi}_k \right)}{\lambda\left( \xi _n \right)}\left( \frac{\lambda\left( \hat{\xi}_k \right)}{\hat{\xi}_k-\xi _n}-\lambda^{'}\left( \hat{\xi}_k \right) \right) +\frac{i\xi _{k}^{3}}{q_{-}^{*}q_{0}^{2}}\delta _{k,n}.
	\end{align*}
	Combining Eq.~(\ref{e36}) with Eq.~(\ref{e37}), we have the explicit expression of the double-pole solution:
	\begin{equation}
		q\left( x,t \right) =\left( \frac{\det \left( \hat{H} \right)}{\det \left( \hat{R} \right)} \right) ^2\frac{\det \left( R \right)}{\det \left( H \right)}\left( 1+\frac{\det \left( R \right)}{\det \left( H \right)} \right) q_- .
	\end{equation}	
		Since $\sigma$ can be unified through a transformation, therefore we will exclusively focus on the case where $\sigma=-1$. 
	
	Case 1: $\sigma=-1,\ \eta=1$. 
	In this case, $\theta$ condition (\ref{e39}) reduces to 
	\begin{equation}
		\frac{q_-^{\ast}}{q_-}=\prod_{n=1}^{N_1}{\frac{\zeta _n^8}{q_0^8}}e^{8i\sum_{k=1}^{N_2}{arg\delta _k}}e^{2i\bar{m}}.
	\end{equation}
	
	$\bullet$ As $N_1=0,\ N_2=1,\ $we let $q_-=1,\ \delta_1=e^{i\theta},\ \theta \in \left( 0,\frac{\pi}{2} \right).\ $Then, from formula (\ref{e40}), we know that $b\left( \delta _1 \right) b\left( -\delta _1 \right) =-1,\ d\left( -\delta_1 \right) =-b^2\left( -\delta_1 \right) d\left( \delta_1 \right)$. By selecting different values, we obtain the bright–dark soliton solution shown in Fig. 6.
	\begin{figure}[H]
		\centering
		\begin{minipage}[c]{0.31\textwidth} 
			\centering
			\includegraphics[height=4cm,width=\textwidth]{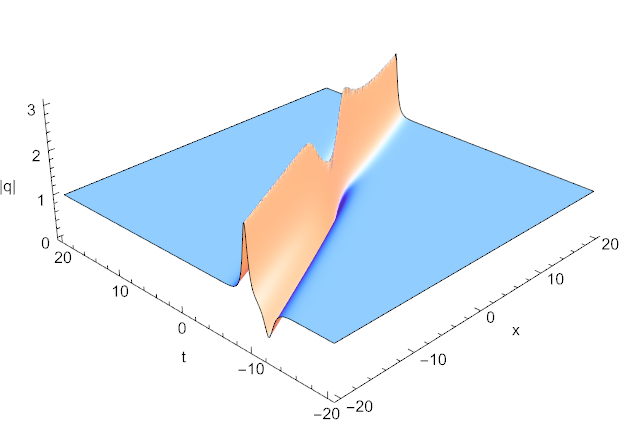} 
			\centerline{(a)}
		\end{minipage}
		\begin{minipage}[c]{0.31\textwidth}
			\centering
			\includegraphics[height=4cm,width=\textwidth]{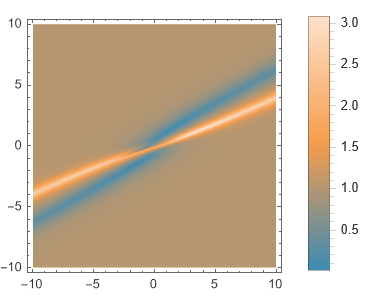}
			\centerline{(b)}
		\end{minipage}
		\begin{minipage}[c]{0.31\textwidth}
			\centering
			\includegraphics[height=4cm,width=\textwidth]{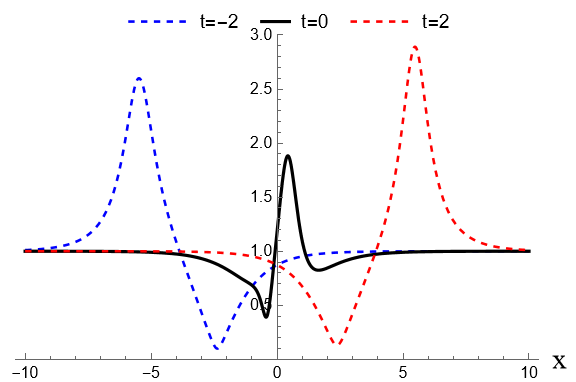}
			\centerline{(c)}
		\end{minipage}
		\caption{Double-pole bright–dark soliton solution with $q_-=1,\ \delta_1=e^{\frac{i\pi}{4}}.\ b\left( \delta _1 \right)=d\left( \delta _1 \right)=1.$ (a) the three-dimensional plot, (b) the two-dimensional density plot, (c) plots of $q(x)$ with three values of t.}
	\end{figure}
		
	Case 2: $\sigma=-1,\ \eta=-1$. 
	In this case, $\theta$ condition (\ref{e39}) reduces to 
	\begin{equation}
		-\frac{q_-^{\ast}}{q_-}=\prod_{n=1}^{N_1}{\frac{\zeta _n^8}{q_0^8}}e^{8i\sum_{k=1}^{N_2}{arg\delta _k}}e^{2i\bar{m}}.
	\end{equation}
	
	$\bullet$ For $N_1=0,\ N_2=1,\ $we let $q_-=1,\ \delta_1=e^{i\theta},\ \theta \in \left( 0,\frac{\pi}{2} \right).\ $Then, from formula (\ref{e40}), we know that $b\left( \delta _1 \right) b\left( -\delta _1 \right) =-1,\ d\left( -\delta _1 \right) =-b^2\left( -\delta _1 \right) d\left( \delta _1 \right)$. By selecting different values, we obtain the bright–dark soliton solution shown in Fig. 7.
	
	\begin{figure}[H]
		\centering
		\begin{minipage}[c]{0.31\textwidth} 
			\centering
			\includegraphics[height=4cm,width=\textwidth]{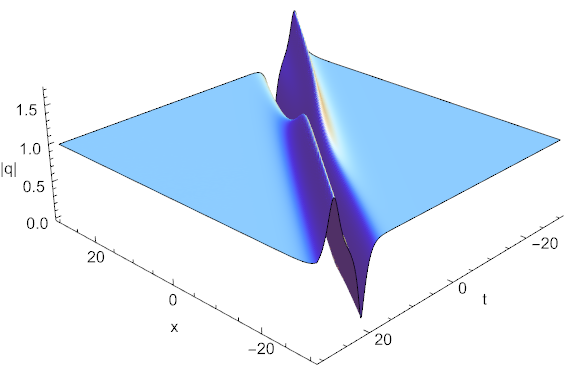} 
			\centerline{(a)}
		\end{minipage}
		\begin{minipage}[c]{0.31\textwidth}
			\centering
			\includegraphics[height=4cm,width=\textwidth]{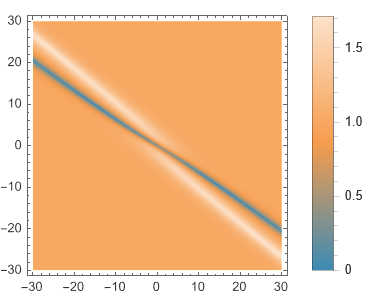}
			\centerline{(b)}
		\end{minipage}
		\begin{minipage}[c]{0.31\textwidth}
			\centering
			\includegraphics[height=4cm,width=\textwidth]{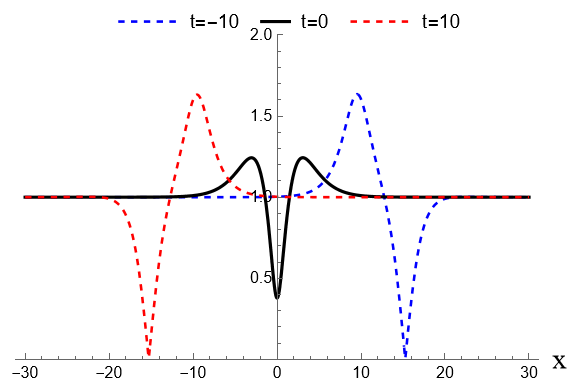}
			\centerline{(c)}
		\end{minipage}
		\caption{Double-pole bright–dark soliton solution with $q_-=1,\ \delta_1=e^{\frac{3i\pi}{8}}.\ b\left( \delta _1 \right)=d\left( \delta _1 \right)=1.$ (a) the three-dimensional plot, (b) the two-dimensional density plot, (c) plots of $q(x)$ with three values of t.}
	\end{figure}
	\section{Conclusions}
	\hspace{0.7cm}In this paper, we have investigated the single-pole and double-pole solutions of the RST-DNLS equation under nonzero boundary conditions via the RH method. The direct scattering problem illustrated the analyticity, symmetries, and asymptotic behaviors of the Jost eigenfunctions and scattering matrix functions. Different from the local DNLS equation, the direct scattering problem of the nonlocal DNLS equation needs to establish new symmertries of the Jost eigenfunctions and discuss more cases. Then, in the inverse scattering problem, we constructed and solved the Riemann-Hilbert problem, which derived the trace formulae, $\theta$ condition, and exact expression of the single-pole and double-pole solutions with the reflectionless potential. Furthermore, we also used graphical simulation to show some representative dark-dark solitons, bright-dark solitons, and breather solutions.
	
	\vspace{5mm}\noindent\textbf{Acknowledgements}
	\\\hspace*{\parindent}We express our sincere thanks to each
	member of our discussion group for their suggestions. This work has been supported by the Fund Program for the Scientific Activities of Selected Returned Overseas Scholars in Shanxi Province under Grant No.20220008, and the Shanxi Province Science Foundation under Grant No.202303021221031.
	
	\vspace{5mm}\noindent\textbf{AUTHOR DECLARATIONS}
	
	\vspace{1mm}\noindent\small\textbf{Conflict of Interest}
	\\\hspace*{\parindent}The authors have no conflicts to disclose.
	
	\vspace{5mm}\noindent\textbf{Author Contributions}
	\\{\bf Xin-Yu Liu}: Formal analysis(equal); Investigation(equal); Methodology(equal); Validation(equal); Writing-original draft (equal). \\
	{\bf Rui Guo}: Formal analysis(equal); Methodology(equal); Supervision(lead); Validation(equal); Writing-review $\&$ editing (lead).
	
	\vspace{5mm}\noindent\textbf{Data Availability}
	\\\hspace*{\parindent}The data that support the findings of this study are available within the article.

\end{document}